\documentclass[pra, twocolumn, floatfix]{revtex4}
\usepackage{graphicx}
\usepackage{color}
\usepackage{amsmath, amsfonts, amssymb, bm}
\begin{document}
\title{Electron-positron pair creation in the superposition of two oscillating
electric field pulses with largely different frequency, duration and relative positioning}
\author{N. Folkerts, J. Putzer, S. Villalba-Ch\'avez, and C. M\"uller}
\address{Institut f\"ur Theoretische Physik I, Heinrich-Heine-Universit\"at D\"usseldorf, Universit\"atsstra{\ss}e 1, 40225 D\"usseldorf, Germany}
\date{\today}
\begin{abstract}
Production of electron-positron pairs in two oscillating strong electric field pulses with largely different frequencies and durations is considered. In a first scenario, the influence of a low-frequency background field on pair production by a short main pulse of high frequency is analyzed. The background field is shown to cause characteristic modifications of the momentum spectra of created particles which, in turn, may be used for imaging of the background pulse. In a second scenario, an ultrashort, relatively weak assisting pulse is superimposed onto a strong main pulse. By studying the dependence of the pair production on the field parameters it is shown that duration and relative position of the ultrashort pulse modify the momentum spectra of produced particles in a distinctive way. Both scenarios enable, moreover, to extract partial information about the time periods when pairs with certain momenta are produced predominantly.
\end{abstract}

\maketitle

\section{Introduction}
In certain classes of electromagnetic fields the quantum vacuum 
can become unstable against production of electron-positron pairs, 
this way transforming pure electromagnetic energy into matter 
\cite{Review1, Review2, Review3, Review4}. In recent years, theoreticians 
have devoted special attention to this subject because dedicated experiments 
on strong-field pair production are being planned at various high-intensity 
laser laboratories worldwide, such as Extreme-Light Infrastructure \cite{ELI}, 
Center for Relativistic Laser Science \cite{CoReLS}, Stanford Linear 
Accelerator Center (SLAC) \cite{E320}, Rutherford Appleton Laboratory \cite{RAL} 
or European X-Ray Free-Electron Laser \cite{LUXE}. 
They are going to open a new era in strong-field physics by
probing uncharted regions of the large parameter space, this way considerably 
extending the so far unique observation of electron-positron pair creation
by multiphoton absorption in strong laser fields at SLAC in the 1990s \cite{SLAC}.

Electric fields periodically alternating in time can serve as simplified 
field configurations to model laser pulses. Specifically, a standing laser 
wave -- formed by the superposition of two counterpropagating laser pulses --
approaches to an oscillating, purely electric field in the vicinity 
of the wave's electric field maxima, where pairs are created predominantly. 
A corresponding approximation is therefore suitable if the 
characteristic pair formation length is much smaller than the laser
wavelength and focusing scale. Pair production from the vacuum induced 
by the presence of an oscillating electric field was first studied in the 1970s 
\cite{Brezin,Popov,Mostepanenko,Gitman}. It was found that various interaction 
regimes exist where the process exhibits qualitatively different behavior. 
They are divided by the ratio of field amplitude $E_0$ and field frequency $\omega$, 
that combine into a dimensionless parameter $\xi = |e| E_0 /(mc\omega)$, 
with electron charge $e<0$, electron mass $m$, and speed of light $c$. While for 
$\xi \ll 1$, the production probability follows a perturbative power-law scaling 
with $E_0$, in the quasistatic case of $\xi \gg 1$ it is distinguised by a manifestly 
nonperturbative exponential dependence on $E_{\rm cr}/E_0$, similarly to Schwinger 
pair production in a constant electric field \cite{Review1, Review2, Review3, Review4}. 
Here, $E_{\rm cr}=m^2c^3/(|e|\hbar)\gg E_0$ denotes 
the critical field strength of QED. Situated in between these asymptotic regimes 
is the nonperturbative domain of intermediate coupling strengths $\xi \sim 1$, 
where analytical treatments of the problem are very difficult.
Noteworthy, a close analogy with strong-field photoionization in intense laser 
fields exists where the corresponding regimes of perturbative multiphoton ionization, 
tunneling ionization and above-threshold ionization are well known.

While the seminal papers \cite{Brezin,Popov,Mostepanenko,Gitman} relied on 
monofrequent electric fields of
infinite temporal extent, the physics of pair production becomes even 
richer when more complex field structures are considered. By accounting 
for finite pulse durations \cite{AdP2004,subcycle,Mocken,GrobeJOSA}, 
different pulse shapes \cite{Kohlfurst2013, Plunien2017, Grobe2019} and field 
polarizations \cite{Blinne, Bauke, Francois2017, Xie2017, Kohlfurst2019}, 
frequency chirps \cite{Dumlu2010, Alkofer2019, Grobe2020}, and spatial inhomogeneties 
\cite{Ruf, Alkofer, Dresden, Schutzhold-inhom, Kohlfurst2020,Kohlfurst2022} 
it has turned out that the production process is very sensitive to the 
precise form of the applied field. 

Particularly interesting phenomena arise when two oscillating fields 
of different frequency are superimposed. When the frequencies are commensurate,
characteristic two-pathway quantum interferences and relative-phase effects arise 
\cite{Fofanov,Krajewska,Brass2020}. Coherent amplifications due to multiple-slit 
interferences in the time domain have also been found in sequences 
of electric field pulses \cite{Dunne2012, modulation, grating, Granz}. 
In bifrequent fields composed of a weak high-frequency and a strong low-frequency
component, vast enhancement of pair production is expected to occur through
the dynamically assisted Schwinger effect \cite{Schutzhold2008, Orthaber, 
Grobe2012, Akal, Opt, slit, Kampfer-survey, KampferEPJD, Torgrimsson, KampferEPJA, Selym-PRD}. 
In case of oscillating electric field pulses, the latter was studied for 
pulses with a {\it common} envelope of flat-top \cite{Akal,Kampfer-survey,
Torgrimsson, KampferEPJA}, Gaussian \cite{Opt,KampferEPJD,KampferEPJA} 
or super-Gaussian \cite{KampferEPJD,KampferEPJA} form, so that both pulses 
act during the same time duration. The pair yield was moreover optimized 
with respect to a time-lag between two Gaussian pulses of different widths \cite{Opt}.

In the present paper, we study electron-positron pair creation in superpositions
of two oscillating electric field pulses. Taking the known phenomenology of the
process in a single electric-field pulse as reference (see, e.g., \cite{Mocken})
we address the question of how the momentum spectra are modified by the 
additional presence of either (i) a low-frequency background field or (ii) 
an ultrashort pulse of very high frequency. The first of these scenarios 
qualitatively resembles the field configuration that is applied for streak
imaging in atomic physics \cite{streaking1,streaking2}; the second scenario is related to
the phenomenon of dynamical assistance and extends a previous study where two
frequency modes of same temporal duration were superimposed \cite{Akal}. The 
guiding objective throughout is to reveal the impact that the positioning of 
both pulses relative to each other exerts on the pair production process. 

Our paper is organized as follows. In Sec.~II we present our computational
approach to calculate the momentum-dependent probabilities for pair creation
in time-varying electric field pulses. In Sec.~III we discuss the scenario
where pairs are produced by a strong electric field pulse of high frequency 
in the presence of a low-frequency background field. The complementary situation,
where an ultrashort pulse is superimposed on a strong main pulse is considered
in Sec.~IV. We summarize our findings in Sec.~V.
Relativistic units with $\hbar = c = 1$ are used.

\section{Computational approach}
Our goal is to investigate pair production in the superposition of two oscillating electric field pulses. We chose the fields to be linearly polarized in $y$-direction. In temporal gauge, the total field $\vec E(t)=-\dot{\vec A}(t)$ can be described by a vector potential of the form $\vec{A}(t) = A(t)\vec{e}_y$, with
\begin{eqnarray}
A(t) = A_1(t-t_1) F_1(t-t_1) + A_2(t-t_2) F_2(t-t_2)\,,
\label{A}
\end{eqnarray}
where $t_j$ denote the starting times of the pulses ($j\in\{1,2\}$), both of which having sinusoidal time dependence
\begin{eqnarray}
A_j(t) = \frac{m\xi_j}{e}\,\sin(\omega_j t)\ .
\end{eqnarray}
The envelope functions $F_j(t)$ have compact support on $[0,T_j]$, with turn-on and turn-off increments of half-cycle duration each and a constant plateau of unit height in between. Here, 
$T_j=\frac{2\pi}{\omega_j}N_j$ denote the pulse durations, with the number $N_j$ of oscillation cycles.
We note that, in the numerical examples considered in Sec.\,III, the shorter of the two pulses will always be fully encompassed within the envelope of the longer pulse. The electric field amplitudes of the modes are $E_j = m\xi_j\omega_j/|e|$.

The pair production probability in a time-dependent electric field can be obtained by solving a coupled system of ordinary differential equations \cite{Mostepanenko, Gitman, Mocken, Akal, KampferEPJD, Hamlet}. We use the following representation that was derived in \cite{Mocken, Hamlet}:
\begin{eqnarray}
\dot{f}(t) &=& \kappa(t)f(t) + \nu(t)g(t)\ ,\nonumber\\
\dot{g}(t) &=& -\nu^*(t)f(t) + \kappa^*(t)g(t)\ ,
\label{system}
\end{eqnarray}
with 
\begin{eqnarray}
\kappa(t) &=& ieA(t)\,\frac{p_y}{p_0}\ ,\nonumber\\
\nu(t) &=& -ieA(t)\,e^{2ip_0 t}\,\left[ \frac{(p_x-ip_y)p_y}{p_0(p_0+m)} + i\, \right]\ .
\label{nu}
\end{eqnarray}
It is obtained from the time-dependent Dirac equation when an ansatz of the form $\psi_{\vec p}(\vec r,t) = f(t)\, \phi_{\vec p}^{(+)}(\vec r,t) + g(t)\, \phi_{\vec p}^{(-)}(\vec r,t)$ is inserted. Here, $\phi_{\vec p}^{(\pm)}\sim e^{i(\vec p\cdot\vec r \mp p_0 t)}$, with $p_0=\sqrt{{\vec p}^{\,2}+m^2}$, denote free Dirac states with momentum $\vec p$ and positive or negative energy. The suitability of this ansatz relies first of all on the fact that the canonical momentum is conserved in a spatially homogeneous external field, according to Noether's theorem. Since the canonical momentum coincides with the kinetic momentum $\vec p$ of a free particle outside the time interval when the field is present, it is possible to treat the invariant subspace spanned by the usual four free Dirac states with momentum $\vec p$ separately. Due to the rotational symmetry of the problem about the field axis, the momentum vector can be parametrized as $\vec p = (p_x,p_y,0)$ with transversal (longitudinal) component $p_x$ ($p_y$). As a consequence, one can find a conserved spin-like operator, which allows to reduce the effective dimensionality of the problem further from four to two basis states \cite{Mocken, Hamlet}.

Accordingly, the time-dependent coefficients $f(t)$ and $g(t)$ describe the occupation amplitudes of a positive-energy and negative-energy state, respectively. The system of differential equations \eqref{system} is solved with the initial conditions $f(0)=0$, $g(0)=1$. At time $T=\mbox{max}\{t_1+T_1, t_2+T_2\}$ when the fields have been switched off, $f(T)$ represents the occupation amplitude of an electron state with momentum $\vec p$, positive energy $p_0$ and certain spin projection. Taking the two possible spin degrees of freedom into account, we obtain the probability for creation of a pair with given momentum as 
\begin{eqnarray}
W(p_x,p_y) = 2\,|f(T)|^2\ .
\label{W}
\end{eqnarray}
Note that the created positron has momentum $-\vec p$, so that the total momentum of each pair vanishes.

From previous studies in a monofrequent electric field with potential $A_1(t)$ on the interval $[0,T_1]$ it is known that the pair production shows characteristic resonances whenever the ratio between the energy gap and the field frequency attains integer values \cite{Popov, Mocken, Bauke, GrobeJOSA}. The energy gap is given by $2q_0$, with the time-averaged particle quasi-energies \cite{q0}
\begin{eqnarray}
q_0(\vec{p}\,) = \frac{1}{T_1}\int_0^{T_1}\sqrt{m^2+p_x^2+[p_y - eA_1(t)]^2}\,dt\ .
\label{epsilon}
\end{eqnarray}
For example, in a monofrequent field with $\xi_1=1$ one obtains $q_0(\vec{0}\,)\approx 1.21m$ for vanishing momenta; the difference as compared with the corresponding field-free energy $p_0=m$ is a result of field dressing. As a consequence, a field frequency of $\omega_1\approx 0.35m$ leads to resonant production of particles at rest by absorption of seven field quanta (``photons''), for instance \cite{Mocken}. To allow for a comparison of our results with this earlier study, we will employ similar frequency values $\omega_1\approx 0.3m$--0.5$m$ for the main pulse. Applying besides the normalized amplitude $\xi_1=1$ places us into the nonperturbative multiphoton regime of pair production. 

We emphasize that such high field frequencies are also considered for reasons of computational feasibility. In this case, however, a purely time-dependent field represents just a simplified model (rather than a close approximation) for the electromagnetic fields of a standing laser wave. Significant differences between pair production in an oscillating electric field and pair production in a standing laser wave are known to arise at frequencies $\gtrsim 0.1m$ from the spatial dependence and magnetic component of the latter \cite{Ruf, Alkofer, Dresden, Schutzhold-inhom, Kohlfurst2020}. Some general,  qualitative features of the impact of the superimposed second field pulse that shall be discussed below may nevertheless be expected to find their counterparts in laser-induced pair production as well.

\section{Superposition of low-frequency background field}
\begin{figure}[b]  
\vspace{-0.25cm}
\begin{center}
\includegraphics[width=0.4\textwidth]{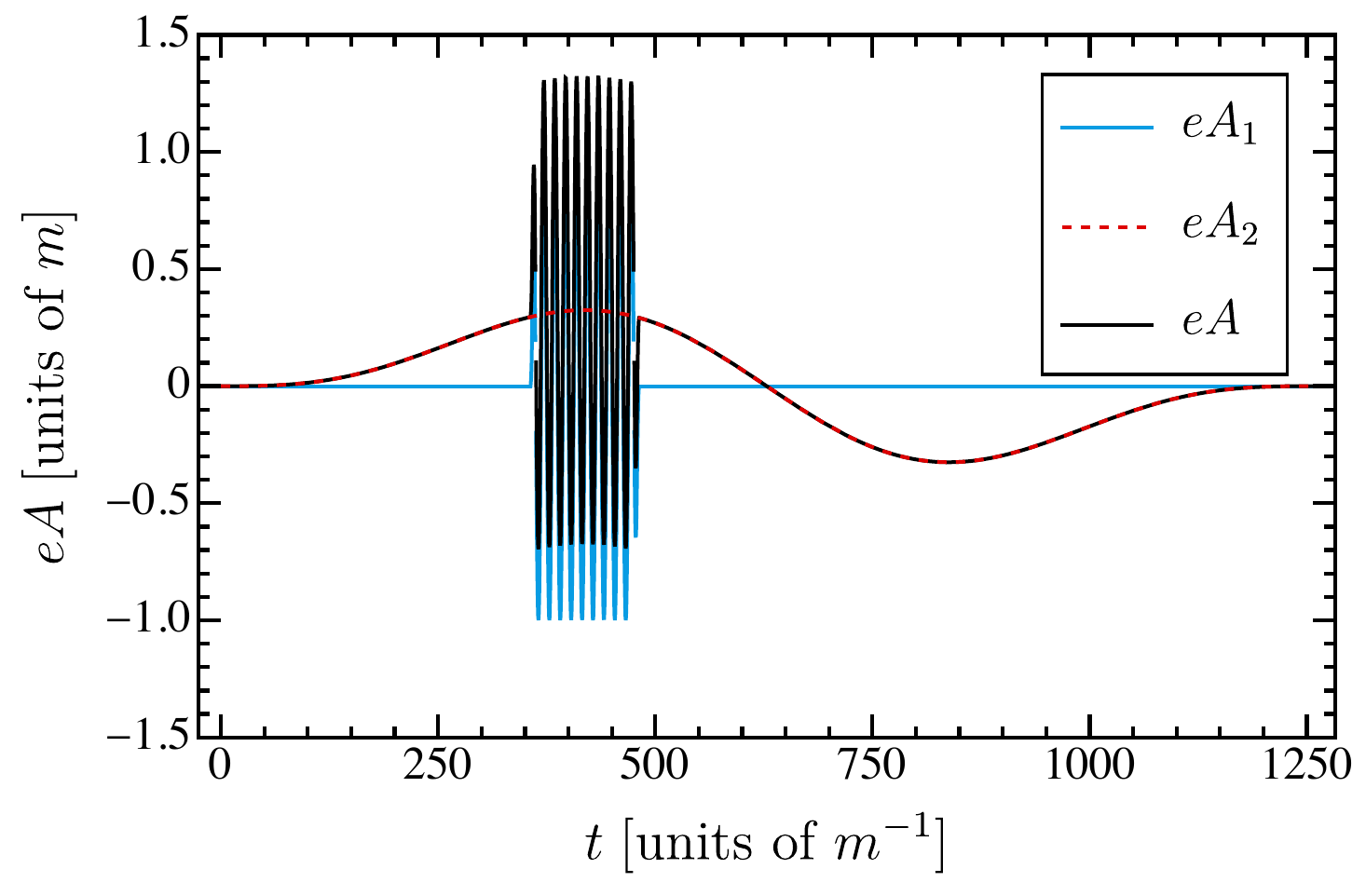}
\end{center}
\vspace{-0.5cm} 
\caption{Vector potentials, multiplied by $e$, of the fast oscillating main pulse $A_1$, the slowly varying background pulse $A_2$, and their superposition $A$, as indicated in the legend. The time delay amounts to $t_1 = \Delta t^{(1)} = \frac{1}{3}T_2-\frac{1}{2}T_1$ here.}
\label{figA1A2-background}
\end{figure}

In our first scenario we consider pair production by a strong electric field pulse of high frequency ($\xi_1=1$, $\omega_1\lesssim m$, $E_1\lesssim E_{\rm cr}$) onto which a background field of very low frequency and moderate field strength ($\xi_2\lesssim \xi_1$, $\omega_2\ll\omega_1$, $E_2\ll E_1$) is superimposed. Our goal is to reveal the impact that such a slow background field can exert on the pair production process. The field structure is described by Eq.~\eqref{A}, with sin$^2$-shaped turn-on and turn-off segments of half a period; see Fig.~\ref{figA1A2-background} for an illustration. In this section, the background field always starts at $t_2=0$; correspondingly, a relative time delay between the pulses is described by the start time $t_1$ of the main pulse.

\begin{figure}[b]  
\begin{center}
\includegraphics[width=0.425\textwidth]{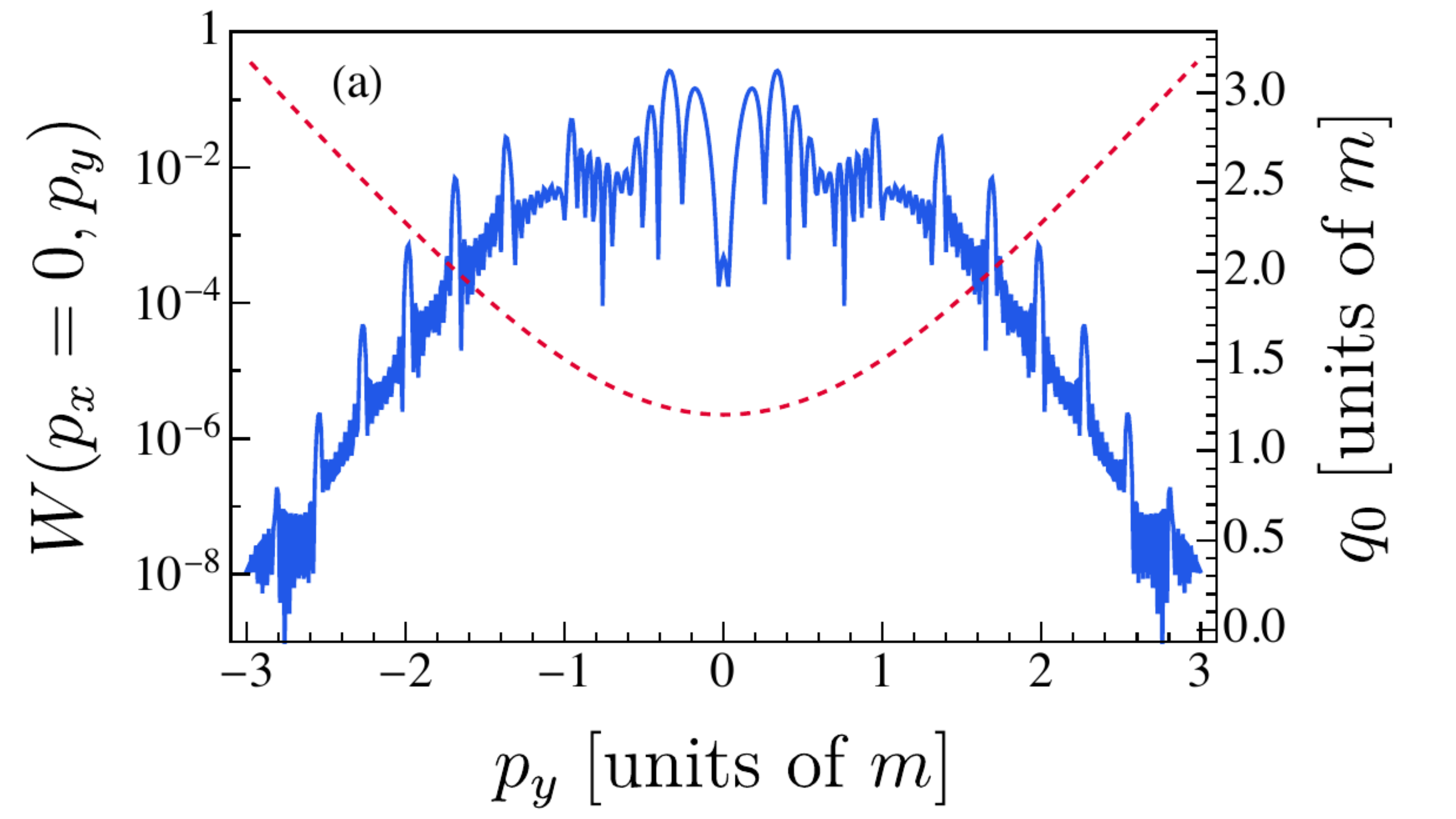}
\includegraphics[width=0.425\textwidth]{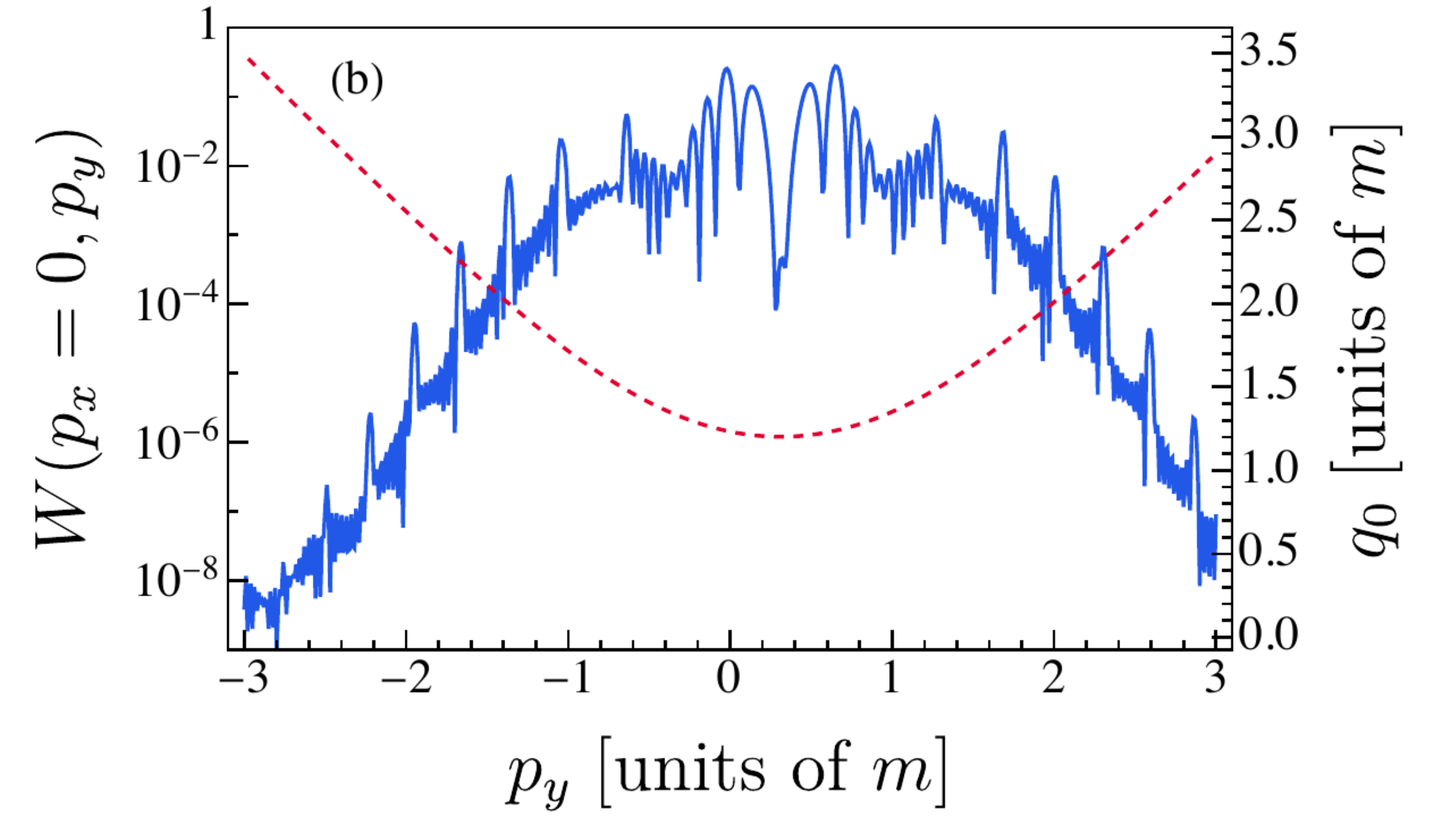}
\includegraphics[width=0.425\textwidth]{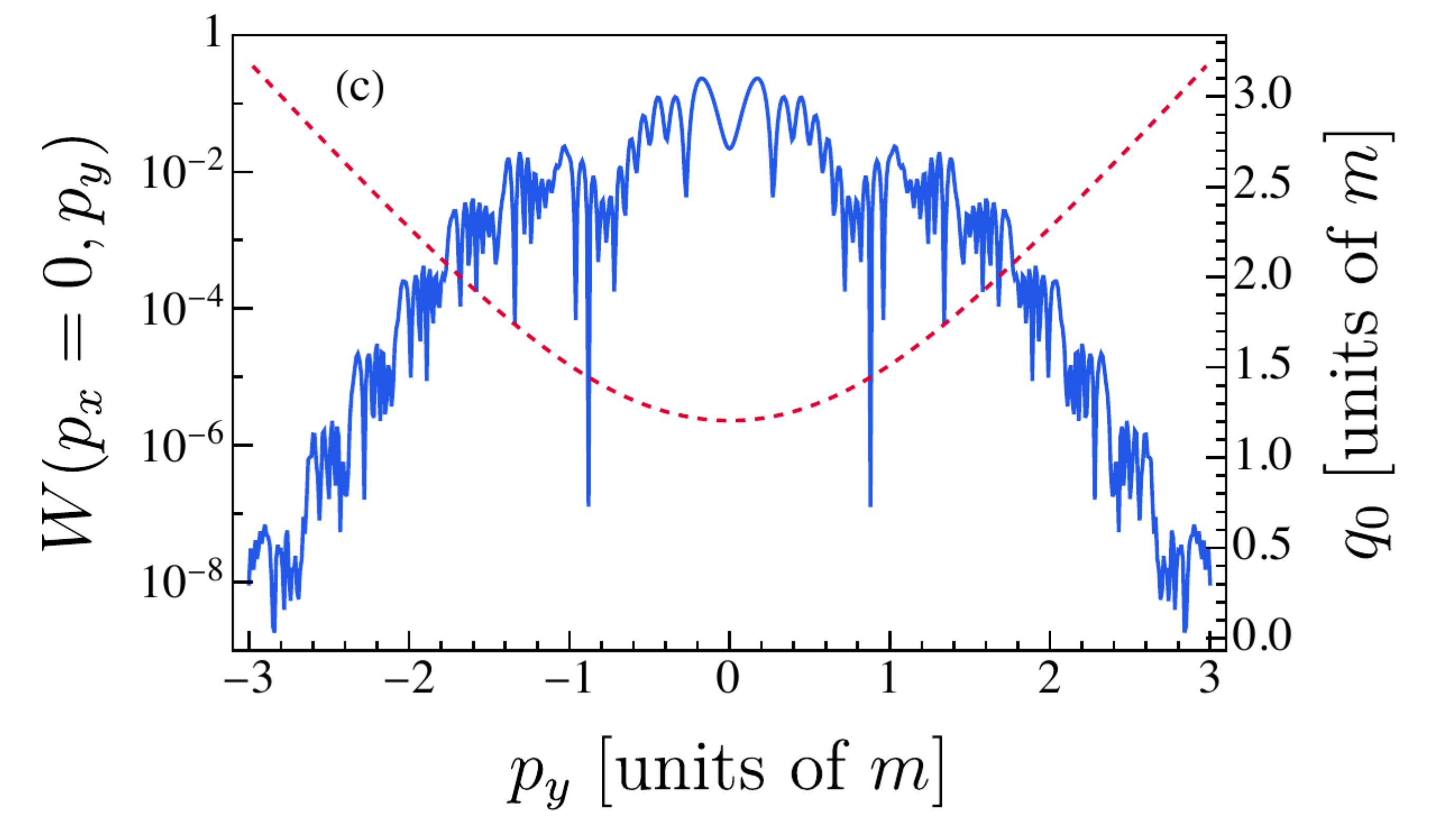}
\includegraphics[width=0.425\textwidth]{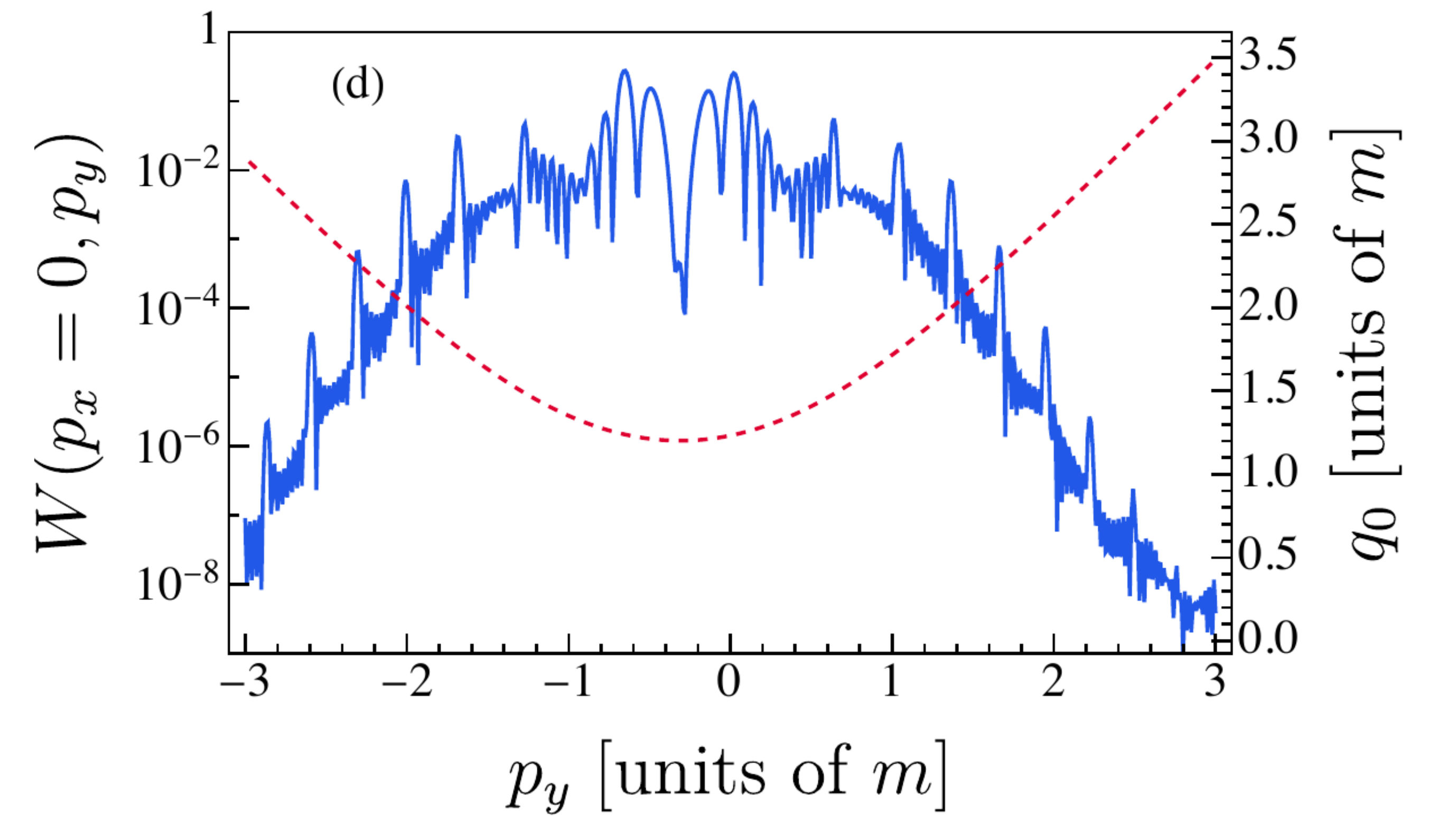}
\end{center}
\vspace{-0.5cm} 
\caption{Longitudinal momentum distributions of created electrons (blue solid lines). Panel (a) refers to a monofrequent electric field with $\xi_1=1$, $\omega_1=0.5m$, $N_1=10$. Panels (b)-(d) show the results for a bifrequent electric field with $\xi_1=1$, $\xi_2=0.5$, $\omega_1=0.5m$, $\omega_2=0.005m$, $N_1=10$, $N_2=1$ and relative time delay $t_1=\Delta t^{(1)}$, $\Delta t^{(2)}$ and $\Delta t^{(3)}$, respectively. The red dashed curves show the corresponding effective energy $q_0(p_y)$. The transverse momentum vanishes, $p_x=0$.}
\label{fig-shift}
\end{figure}

For reference, Fig.~\ref{fig-shift}\,(a) shows the longitudinal momentum distribution of the electrons created by a single, monofrequent electric field with $\xi_1=1$, $\omega_1=0.5m$ and $N_1=10$. The total duration of the field thus amounts to $T_1 = \frac{2\pi N_1}{\omega_1}\approx 125.7\,m^{-1}$. A regular structure of resonance peaks can be seen that correspond to the absorption of an integer number of field quanta. Panels (b)-(d) show how the momentum distribution changes when the pair production occurs in the additional presence of a background field with parameters $\xi_2=0.5$, $\omega_2=0.005m$ and $N_2=1$, implying $T_2 = \frac{2\pi N_2}{\omega_2} = 10T_1$. (Note that the background pulse $A_2(t)$ does not comprise a plateau region since $N_2=1$.) The panels refer to different relative positions of the main pulse, whose delay $t_1$ with respect to the background pulse is chosen as $\Delta t^{(k)} = \frac{k+1}{6}T_2-\frac{1}{2}T_1$ with $k\in\{1,2,3\}$. Accordingly, in Fig.~\ref{fig-shift}\,(b) the main pulse is centered around the maximum of $eA_2(t)$ (as illustrated in Fig.~\ref{figA1A2-background}), in panel (c) it lies in the middle of the background field where $eA_2(t)\approx 0$, and in panel (d) it is centered around the minimum of $eA_2(t)$. While the distribution in panel (c) is symmetric under $p_y\to -p_y$, one observes a clear shift of the spectrum to the right [left] in panel (b) [panel (d)]. This shift can be understood by noting that the kinetic momentum $p_{\rm kin}(t_c)$ of the electron 'at the moment $t_c$ of creation' is related to the canonical momentum $p_y$ by $p_{\rm kin}(t_c) = p_y - eA(t_c)$. In the situations of Fig.~\ref{fig-shift}\,(b) and (d) the vector potential of the background field $A_2$ is nearly constant during the interval when the main pulse is present. Its mean value---multiplied by $e$---during the time interval $[t_1,t_1+T_1]$ amounts to $\langle eA_2\rangle_{T_1}\approx 0.3m$ in the situation of Fig.~\ref{fig-shift}\,(b) and has the opposite sign in Fig.~\ref{fig-shift}\,(d). These values fit very well to the horizontal shifts of the corresponding momentum distributions, because:

During the time interval $T_1\ll T_2$ of $A_1(t)$ when pair production mainly occurs, the effective energy changes by virtue of $p_y-eA_1(t)\to p_y - \langle eA_2\rangle_{T_1} - eA_1(t)$. If a resonance occurs at $p_y=p_r$ in the absence of $A_2(t)$, it will shift to $p_y=p_r + \langle eA_2\rangle_{T_1}$ in the presence of $A_2(t)$. For $\langle eA_2\rangle_{T_1}>0$ like in Fig.~\ref{fig-shift}\,(b), the peak will thus be shifted to the right. Conversely, in Fig.~\ref{fig-shift}\,(d) where $\langle eA_2\rangle_{T_1}<0$, the peak is shifted to the left. Accordingly, the whole $q_0$ curve -- evaluated during the time intervall $T_1$ when both fields are present -- experiences this shift. 

A background field $A_2(t)$ of rather low frequency and amplitude can, thus, modify the momentum distribution of created pairs in a characteristic manner. This pronounced influence is interesting because $A_2(t)$ acting alone would produce almost no pairs at all: the corresponding pair production probability is $\lesssim 10^{-13}$ throughout the considered $p_y$-range. Apart from the shifting effect, the momentum distributions and resonance peak structures in Figs.~\ref{fig-shift}\,(b) and (d) closely resemble the monofrequent result in Fig.~\ref{fig-shift}\,(a).

While no shifting effect arises when the main pulse $A_1(t)$ is located in the middle of the background field [see Fig.~\ref{fig-shift}\,(c)], the 'monofrequent' resonance peak at $p_y=0$ is split here as well. Moreover, the other resonance peaks are much less pronounced, which can be related to the significant variation of $A_2(t)$ while the main pulse is on. As a consequence, the momenta at the moment of creation are to some extent smeared out, leading to a broadening and lowering of the resonance peaks. 

As general conclusion from Fig.~\ref{fig-shift} may be drawn that the background field $A_2$ leads to a redistribution of the particle momenta created by the strong main pulse $A_1$, while keeping the total proability basically conserved.

\begin{figure}[b]
\vspace{-0.25cm}
\begin{center}
\includegraphics[width=0.4\textwidth]{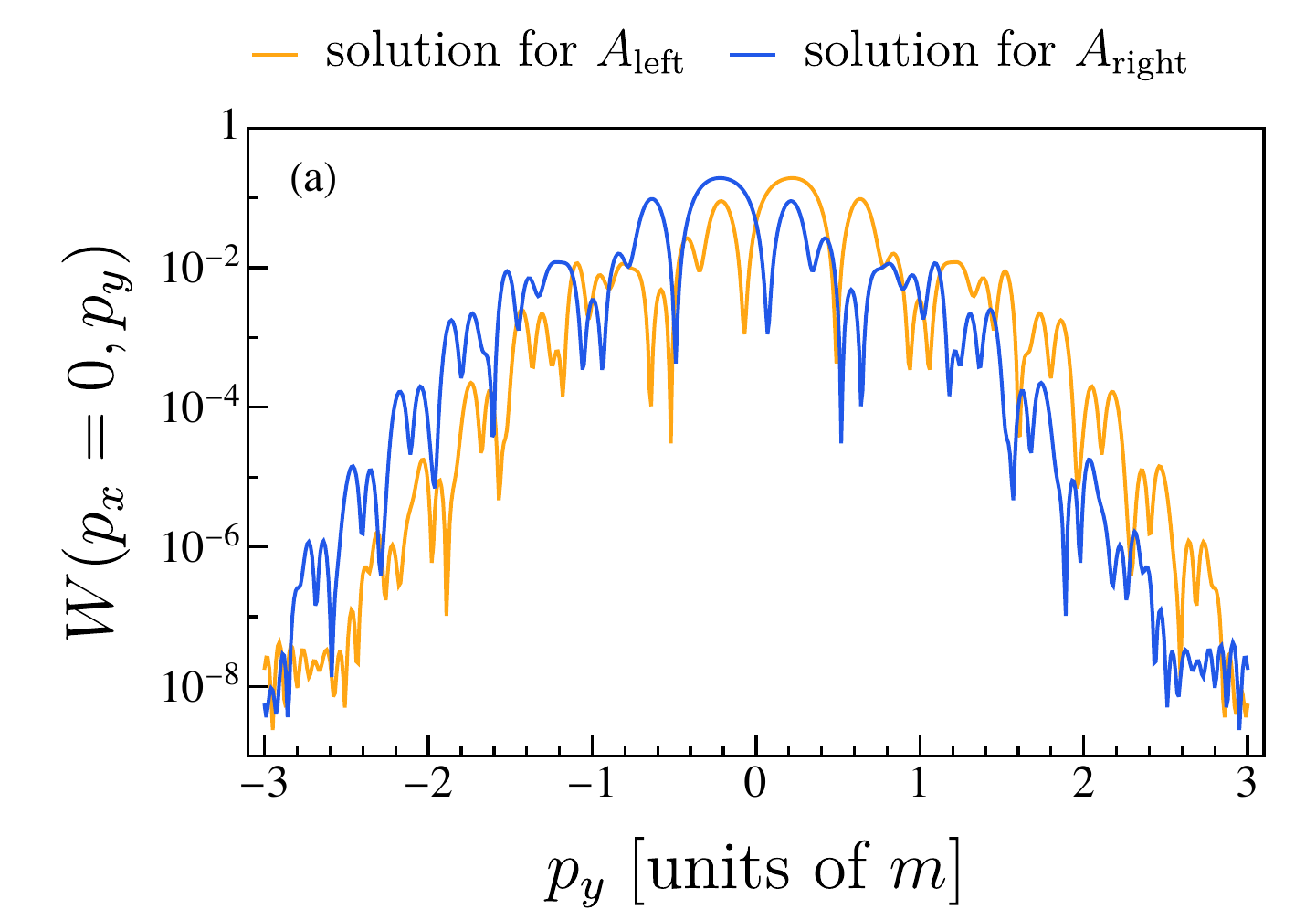}
\includegraphics[width=0.4\textwidth]{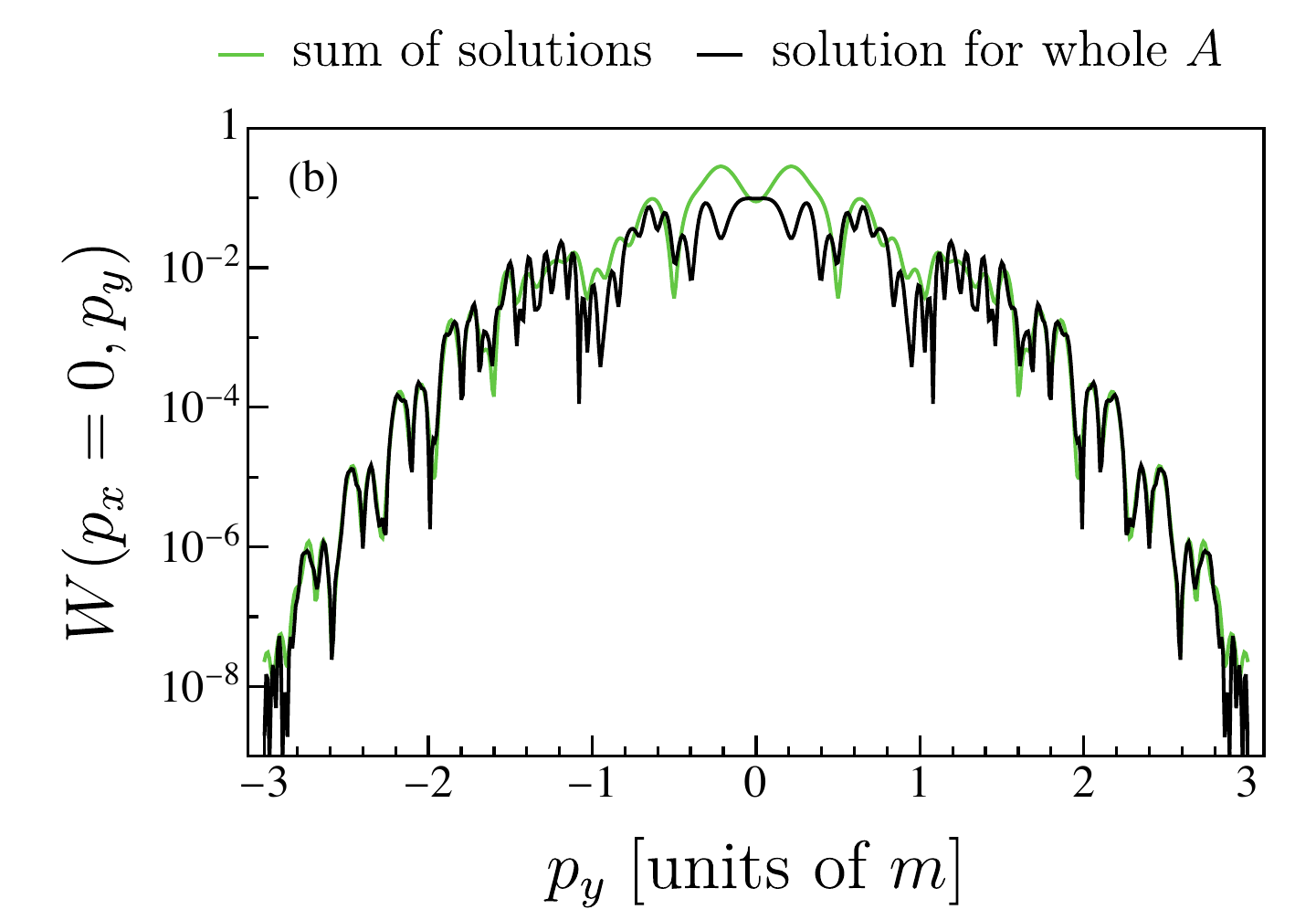}
\end{center}
\vspace{-0.5cm} 
\caption{Longitudinal momentum distribution of electrons created in a bifrequent electric field with 
$\xi_1=1$, $\xi_2=0.5$, $\omega_1=0.5m$, $\omega_2=0.005m$, $N_1=8$, $N_2=1$ and relative time delay $t_1=\Delta t^{(2)}$, so that the main pulse is located in the middle of the background pulse. The yellow [blue] solid line in panel (a) shows the resulting momentum spectrum when -- artificially -- only the left [right] half of the main field acts, as indicated. Panel (b) displays the sum of these separate distributions [green line] and the momentum spectrum that the complete, unsplit field creates [black line], as indicated in the legend.}
\label{fig-split}
\end{figure}

We note that an impact of the precise form of the underlying vector potential on the shape of the resulting momentum distributions of particles has also been found for the nonlinear Bethe-Heitler process in a bichromatic laser field, where characteristic effects of the relative phase between the field modes arise \cite{Krajewska}. For pair creation in a bifrequent oscillating electric field, the relative-phase dependence has recently been analyzed \cite{Brass2020}. The relevance of the carrier-envelope phase for the momentum spectra of pairs produced in ultrashort time-dependent electric-field pulses was demonstrated in \cite{subcycle}.

Before we move on, a comment is in order. The quantity $t_c$ used in our discussion is not sharply defined since pair production is a genuinly quantum process. Nevertheless, on the slow time scale of the background pulse $A_2(t)$ the 'moment of creation' is quite well determined, because the pair production occurs by far predominantly during the short time span during which the main pulse $A_1(t)$ is present, so that $t_c\approx t_1+\frac{1}{2}T_1$. An analogous concept is used in strong-field photoionization \cite{streaking1, streaking2}. We note, moreover, that the time {\it instant} $t_c$ is not to be confused with the 'formation time' which is often considered to describe the chracteristic time {\it duration} for formation of a pair from vaccum in a strong field \cite{Review3}. 

A distinct influence of the background potential has been revealed in Fig.~\ref{fig-shift}\,(b) and (d) when the latter is almost constant during the pair production by the main pulse, whereas the influence was less distinct in the symmetric situation of Fig.~\ref{fig-shift}\,(c). The latter field configuration still raises a very interesting question, though: Since the sign of $eA_2(t)$ during the pair production process matters, one might expect that the spectrum of particles created during the first half of the main pulse (when $eA_2(t)>0$) differs from the spectrum that results from its second half (when $eA_2(t)<0$). To test this hypothesis we have artificially split the main pulse in two halves and calculated the momentum distribution that emerges from each half separately \cite{splitting}. In this calculation, we have used $N_1=8$, $\omega_2=0.02m$ and otherwise the same parameters as before. 

Figure~\ref{fig-split} shows our corresponding results. Panel (a) depicts the separate momentum distributions stemming from the first and second half of the main pulse, respectively. One can clearly see that the first (second) half of the main pulse produces electrons predominantly with positive (negative) momentum values. Panel (b) shows the sum of these two curves, together with the 'full' momentum distribution that results from an unsplit main pulse. One can see that in the outer wings (i.e. for $|p_y|\gtrsim 1.5m$) the sum of the partial distributions matches the full result very well. Thus, the presence of the background field allows us to obtain some information on the typical times of pair production: Electrons with $p_y\gtrsim 1.5m$ are mainly created during the first half of the main pulse, whereas electrons with $p_y\lesssim -1.5m$ mainly emerge during the second half. As the figure shows, this intriguing conclusion still holds to a good approximation for $|p_y|\gtrsim 0.5m$. However, for smaller momenta such a clear timing information cannot be extracted as the sum curve and the full curve differ quite strongly. These momenta are created by both the first half and the second half of the main pulse, so that no clear 'time stamp' can be given. We note in this context that, in general, the appearance of differences between the two curves in Fig.~\ref{fig-split}(b) is to be expected, especially in view of the non-Markovian nature of pair production \cite{non-Markovian}. The corresponding effects might be particularly pronounced around $p_y\approx 0$, as there the pair yields are high.

The characteristic impact of a slow background field on the momenta of electrons promoted into the continuum by a short field pulse of high frequency is exploited in atomic physics for streak imaging \cite{streaking1, streaking2}. In this method, atoms are photoionized by a very short (typically attosecond) laser pulse in the presence of an additional low-frequency (typically femtosecond) laser field. By recording the photoelectron momentum spectrum when the relative delay between the two fields is varied, the shape of the background field can be measured in experiment very accurately, resolving its variation on a femtosecond time scale \cite{streaking2}.

\begin{figure}[b]  
\vspace{-0.25cm}
\begin{center}
\includegraphics[width=0.48\textwidth]{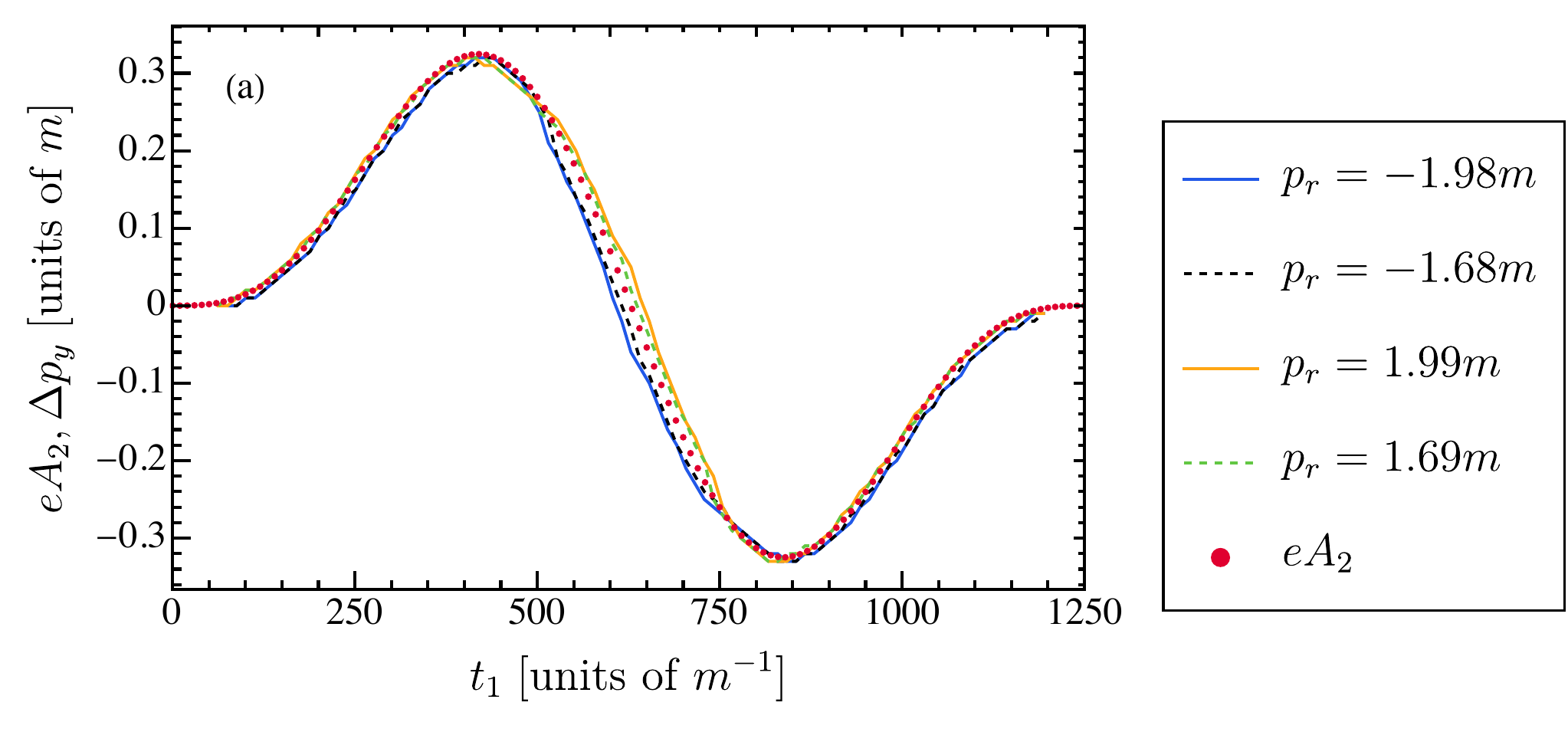}
\includegraphics[width=0.48\textwidth]{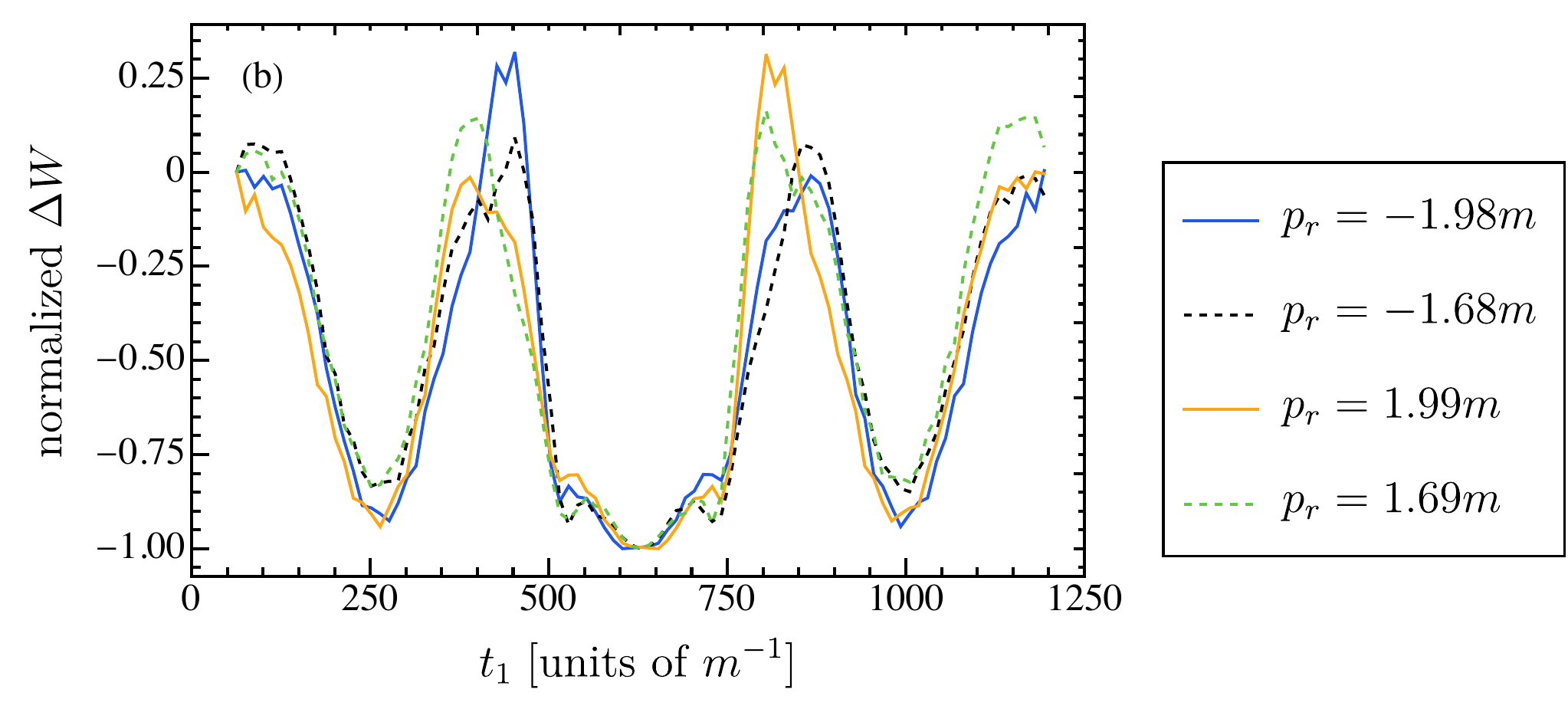}
\end{center}
\vspace{-0.5cm} 
\caption{Panel (a) shows the reconstructed vector potential of the background field that is obtained by tracking the shifts of four selected maxima in the momentum distribution, as indicated. In addition, the  exact curve of $eA_2(t)$ is shown by the circles. Panel (b) displays the normalized heights of these four maxima, as function of the time delay $t_1$.}
\label{fig-imaging}
\end{figure}

In our situation, we can accomplish the same by following selected resonance peaks while varying the time delay $t_1$. As we saw in Fig.~\ref{fig-shift} the peak positions depend sensitively on $A_2(t_c)$. Thus, by monitoring the position of a certain peak as function of $t_1\approx t_c-\frac{1}{2}T_1$, the shape of the background potential can be reconstructed. The outcome of this procedure is shown in Fig.~\ref{fig-imaging}\,(a) for four selected peaks appearing in Fig.~\ref{fig-shift}\,(a). It turns out that the reconstruction works remarkably well \cite{mean-p}.

As we already saw in Fig.~\ref{fig-shift}, the height of the resonance peaks does not remain constant when the pulse delay $t_1$ is varied. This dependence is illustrated in Fig.~\ref{fig-imaging}\,(b) for the same four resonance peaks. The ordinate shows the change of the peak heights relative to their values at $t_1=0$. (Note that the kinks in the curves are not physical but due to small numerical and reading inaccuracies while tracking the moving peaks; the data have been generated along an equidistant $t_1$-grid with 91 points.) One can see that the peak heights are maximal when the main pulse is located in regions where the slope of the background field is small, whereas they are minimal when this slope is large. As we argued in the context of Fig.~\ref{fig-shift}\,(c), a strongly varying background potential leads to a broadening and lowering of the resonances. However, we have checked that the integral resonance strength -- given by the product of peak height and peak width -- stays practically constant. This constancy can explain the shape of the curves in Fig.~\ref{fig-imaging}\,(b): When the main pulse is active during a time intervall where $A_2$ is nearly flat, the resulting resonances are very pronounced, with large height and small width; conversely, when $A_1$ is located in a region where $A_2$ varies strongly, lower and broader peaks result.

Concluding this section we note that the concept of streaking in atomic physics can also be utilized to gain information on the short main pulse \cite{streaking2}. It has been proposed theoretically by considering nonlinear Breit-Wheeler pair production \cite{SHEEP} that this method could in principle be transfered to the relativistic regime, as well, enabling the characterization of ultrashort gamma-ray pulses. We furthermore point out that timing information can generally be extracted from pair creation processes within the theoretical framework of computational quantum field theory (see, e.g., \cite{Grobe-timing}). 

\section{Superposition of ultrashort weak pulse of very high frequency}

\begin{figure}[b]  
\vspace{-0.25cm}
\begin{center}
\includegraphics[width=0.36\textwidth]{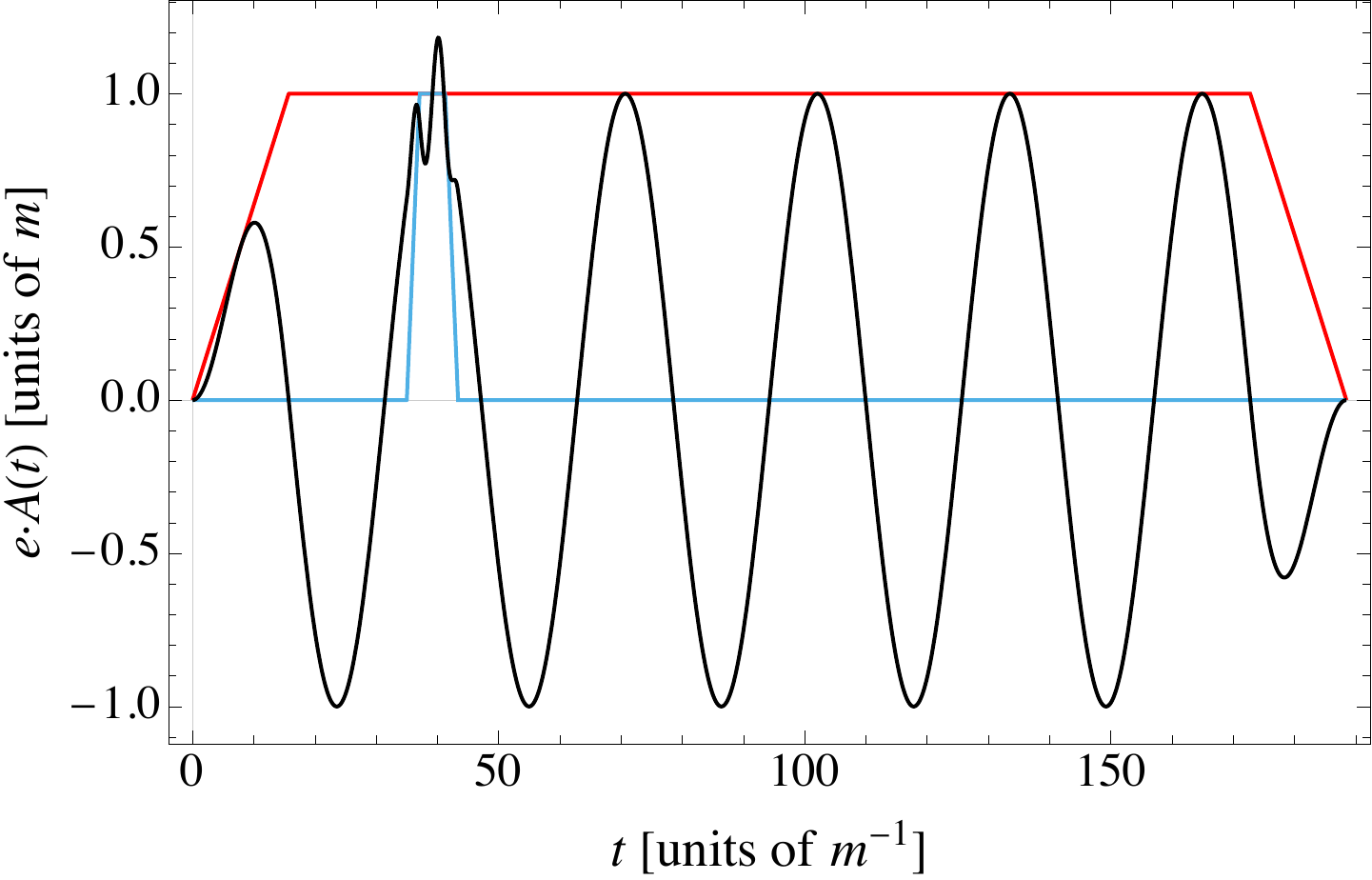}
\end{center}
\vspace{-0.5cm}
\caption{Total vector potential [black line], multiplied by $e$, resulting when a weak ultrashort pulse [whose envelope is shown by the blue (light gray) line] is superimposed with some time delay $t_2$ onto a substantially longer and stronger pulse of lower frequency [with envelope as red (gray) line]. The field parameters are $\xi_1=1$, $\xi_2=0.2$, $\omega_1=0.2m$, $\omega_2=1.5m$, $N_1=5$, $N_2=2$, and $t_2=35m^{-1}$.}
\label{figA1A2-ultrashort}
\end{figure}

In our second scenario we consider a complementary situation: The pair production occurs in the presence of a strong electric field pulse ($\xi_1\approx 1$, $\omega_1\lesssim m$) onto which an ultrashort pulse of moderate intensity and very high frequency is superimposed ($\xi_2\ll 1$, $m\lesssim\omega_2<2m$), as illustrated in Fig.~\ref{figA1A2-ultrashort}. Our goal is to reveal how the pair production is influenced by the duration of the second pulse and its position relative to the first pulse. In this section, the strong pulse will always start at $t_1=0$; a time delay between the pulses is thus described by the start time of the weak ultrashort pulse $t_2$.

\begin{figure}[b]  
\begin{center}
\includegraphics[width=0.36\textwidth]{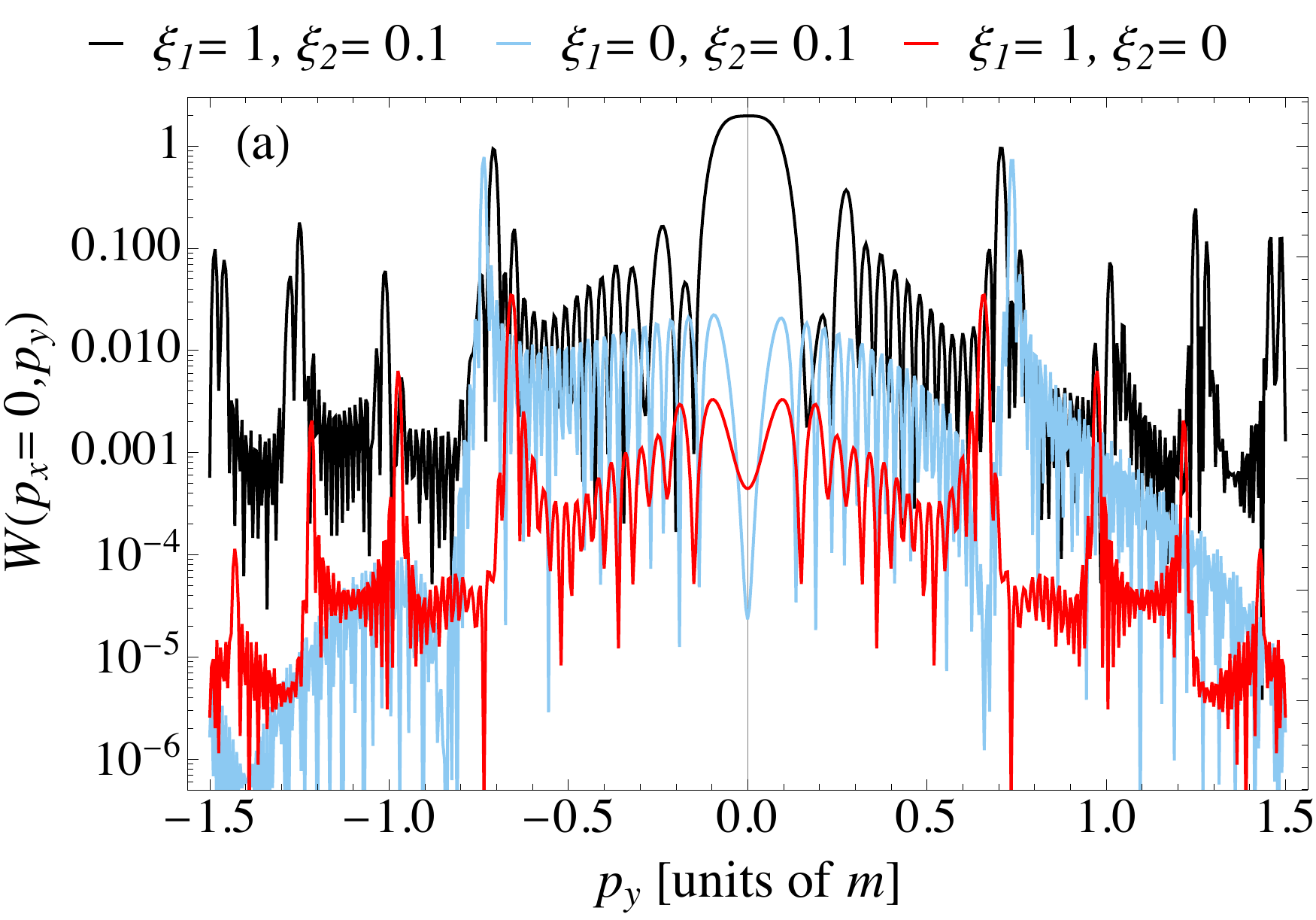}
\includegraphics[width=0.36\textwidth]{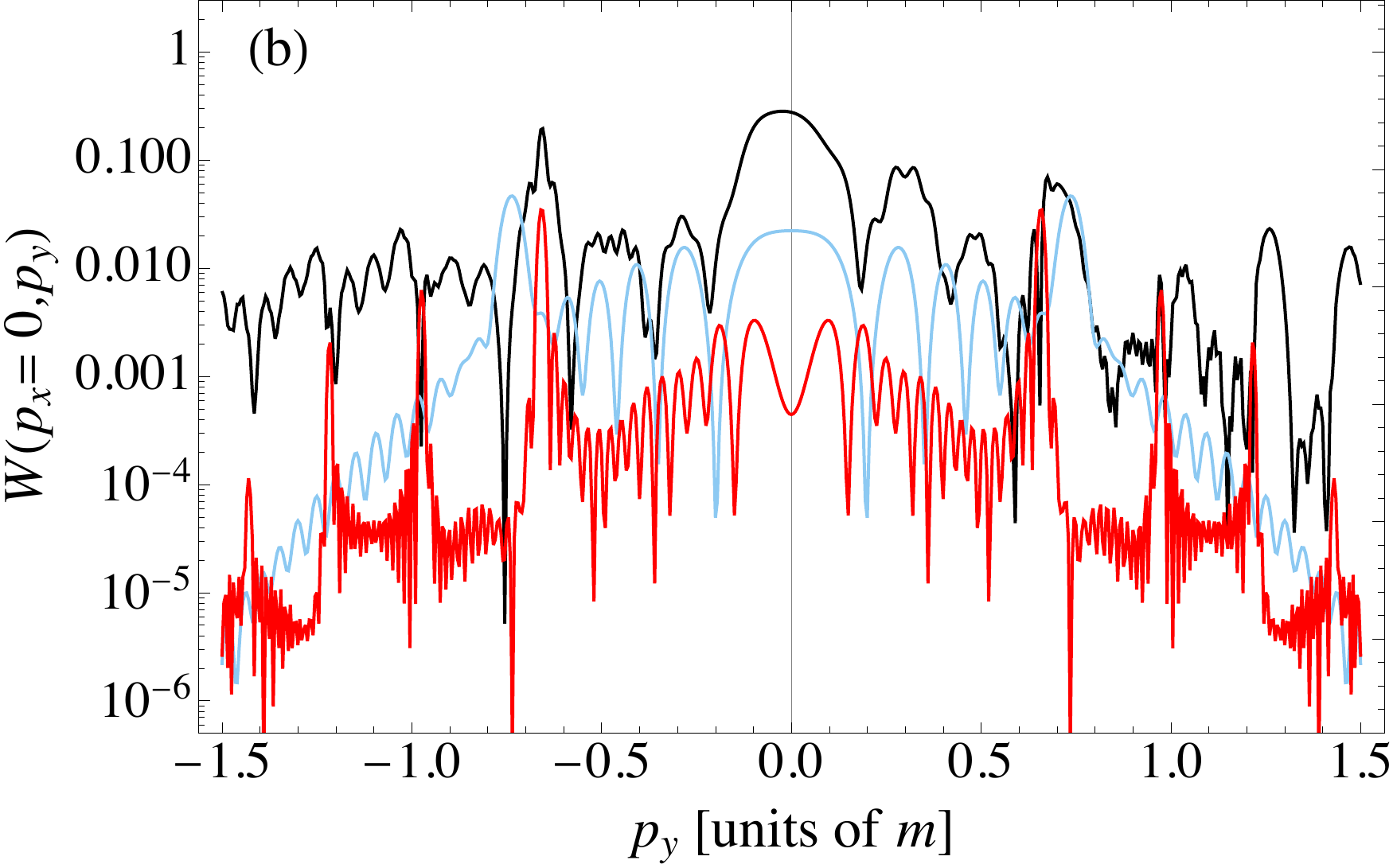}
\includegraphics[width=0.36\textwidth]{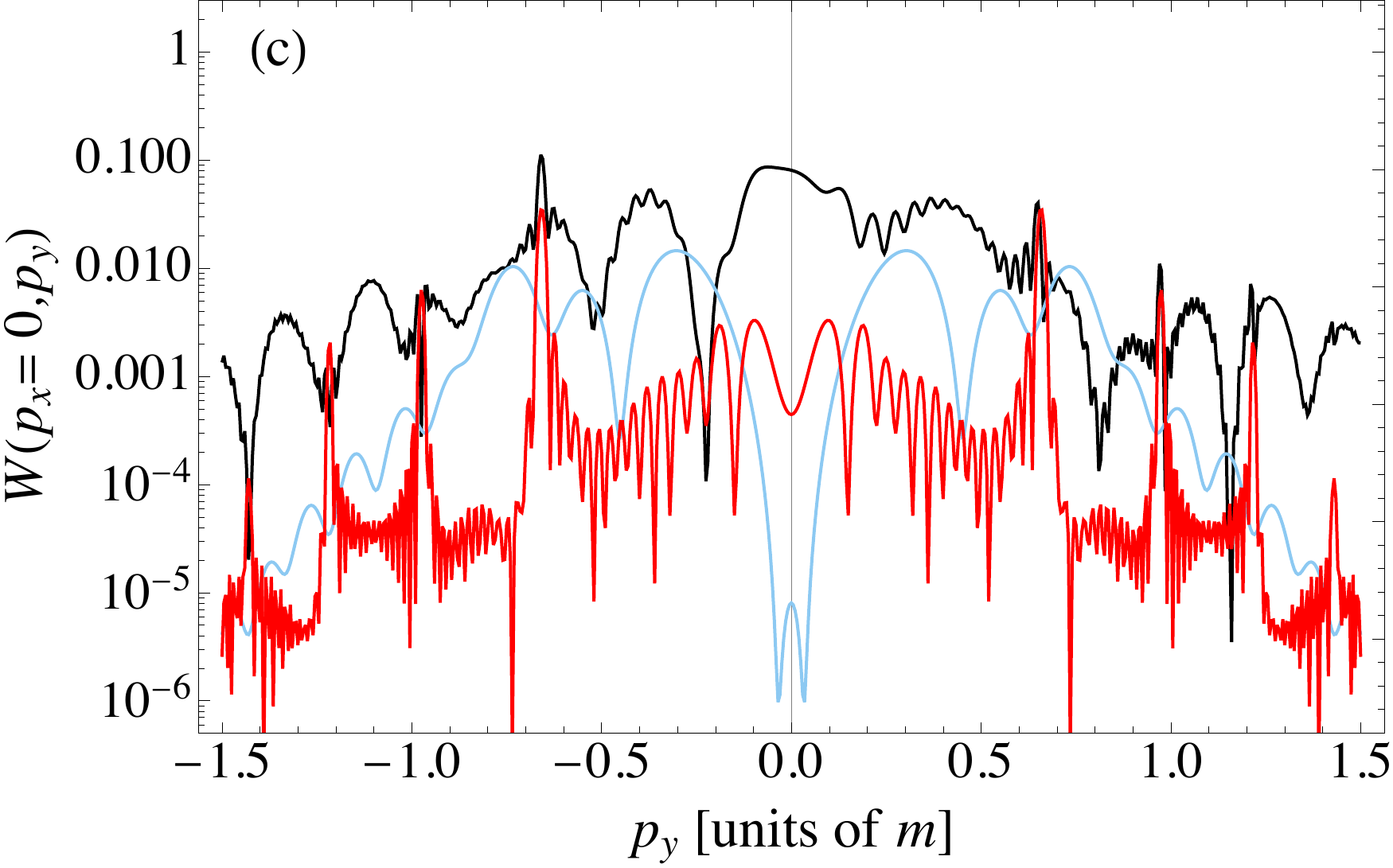}
\includegraphics[width=0.36\textwidth]{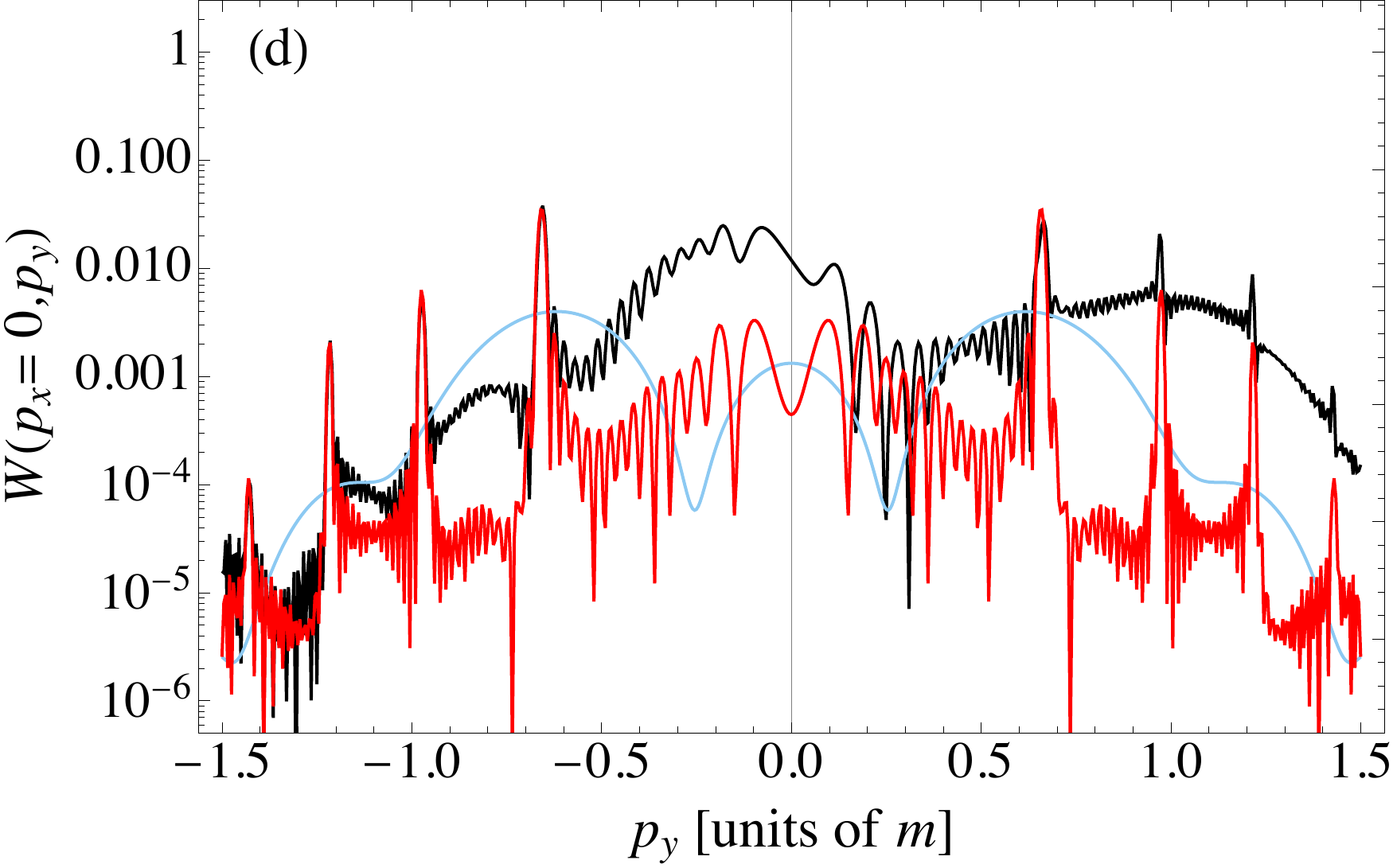}
\end{center}
\vspace{-0.5cm} 
\caption{Longitudinal momentum distributions of electrons (black solid lines) created in a bifrequent electric field with $\xi_1=1$, $\xi_2=0.1$, $\omega_1=0.3m$, $\omega_2=1.24385m$, $N_1=17$, and (a) $N_2=70.5$, (b) $N_2=16$, (c) $N_2=7$ and (d) $N_2=2$, respectively. Both pulses start at the same time, $t_1=t_2=0$. The red (gray) solid lines show the corresponding spectra when solely the main pulse $A_1(t)$ is switched on, whereas the blue (light gray) curves refer to the case when only the assisting ultrashort pulse $A_2(t)$ is active.}
\label{fig-N2varies}
\end{figure}

The situation when both pulses have the same duration has been studied in Ref.~\cite{Akal}; it serves as our reference. The corresponding longitudinal momentum spectrum of the created electrons (with vanishing transverse momentum) is shown in Fig.~\ref{fig-N2varies}\,(a) for $\xi_1=1$, $\omega_1=0.3m$ and $N_1=16$ for the first pulse and $\xi_2=0.1$, $\omega_2=1.24385m$ and $N_2=70.5$ for the second pulse, corresponding to a common pulse duration of $\tau_1=\tau_2 \approx 356m^{-1}$. We note that the calculations in this section have been carried out with trapezoidal envelope functions, possessing linear turn-on and turn-off phases of half a field cycle \cite{sign}. The momentum spectrum for the combined field (black curve) shows distinct resonance peaks that are associated with energy absorption of $n_1\omega_1+n_2\omega_2$ for integer values of $n_1$ and $n_2$. In particular, for the chosen field parameters, a pronounced peak appears at the center around $p_y=0$. Moreover, the height of the spectrum is in general strongly enhanced as compared with the outcomes from each pulse separately, as shown by the red (gray) and blue (light gray) curves, respectively.

In panels (b)-(d) the duration of the second pulse is stepwise reduced to $N_2=16$, $N_2=7$ and $N_2=2$, respectively, with both pulses starting together at time $t_1=t_2=0$. We see that, already for $N_2=16$, the resonance peaks become substantially broader and smear out when $N_2$ is further decreased. This effect can be attributed to the frequency spectrum of the ultrashort pulse whose structuring becomes wider and less oscillatory, as is also reflected by the shape of the momentum spectra resulting from the assisting pulse $A_2$ alone.

While the momentum distribution in Fig.~\ref{fig-N2varies}\,(a) is almost symmetric under $p_y\to -p_y$ (implying that electrons and positrons have practically the same distribution), this symmetry is broken more and more strongly in panels (b)-(d), leading to a pronounced left-right asymmetry that also affects the central peak. In panels (b) and (c) there is still quite a strong overall enhancement of the momentum distribution due to the effect of dynamical assistance by the weak pulse. 
For the shortest pulse duration of $N_2=2$ in panel (d), where $T_2$ has fallen below the duration of a single oscillation cycle of the strong field, there are still regions of very substantial enhancement in the combined fields. But there are now also regions of almost no enhancement, where the probability in the combined fields closely approaches the outcome resulting from the main pulse alone (e.g. for $p_y\lesssim -m$). Thus, by superimposing an ultrashort assisting pulse enhanced production of particles in certain momentum domains is favored. 

\begin{figure}[t]  
\begin{center}
\includegraphics[width=0.4\textwidth]{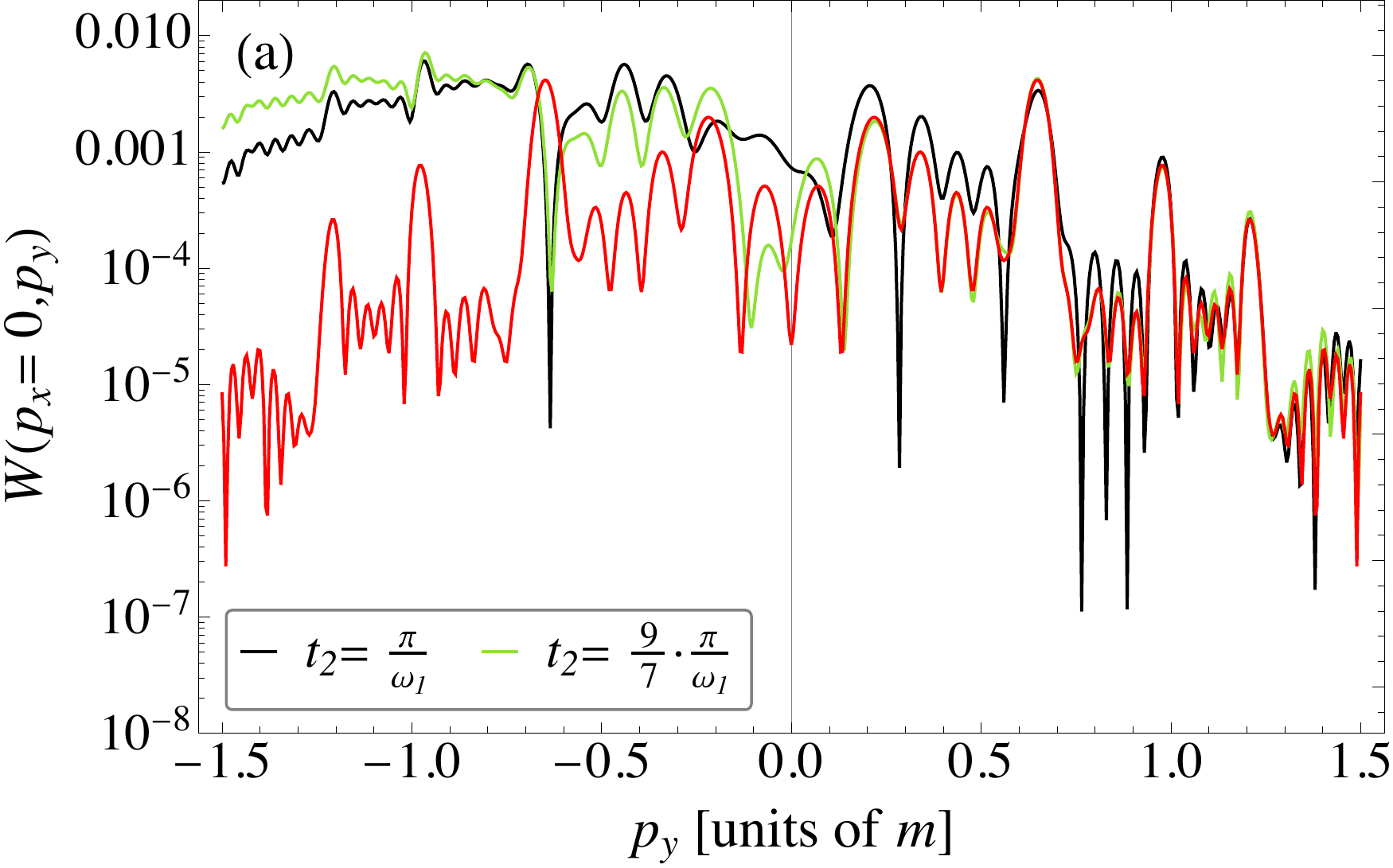}
\includegraphics[width=0.4\textwidth]{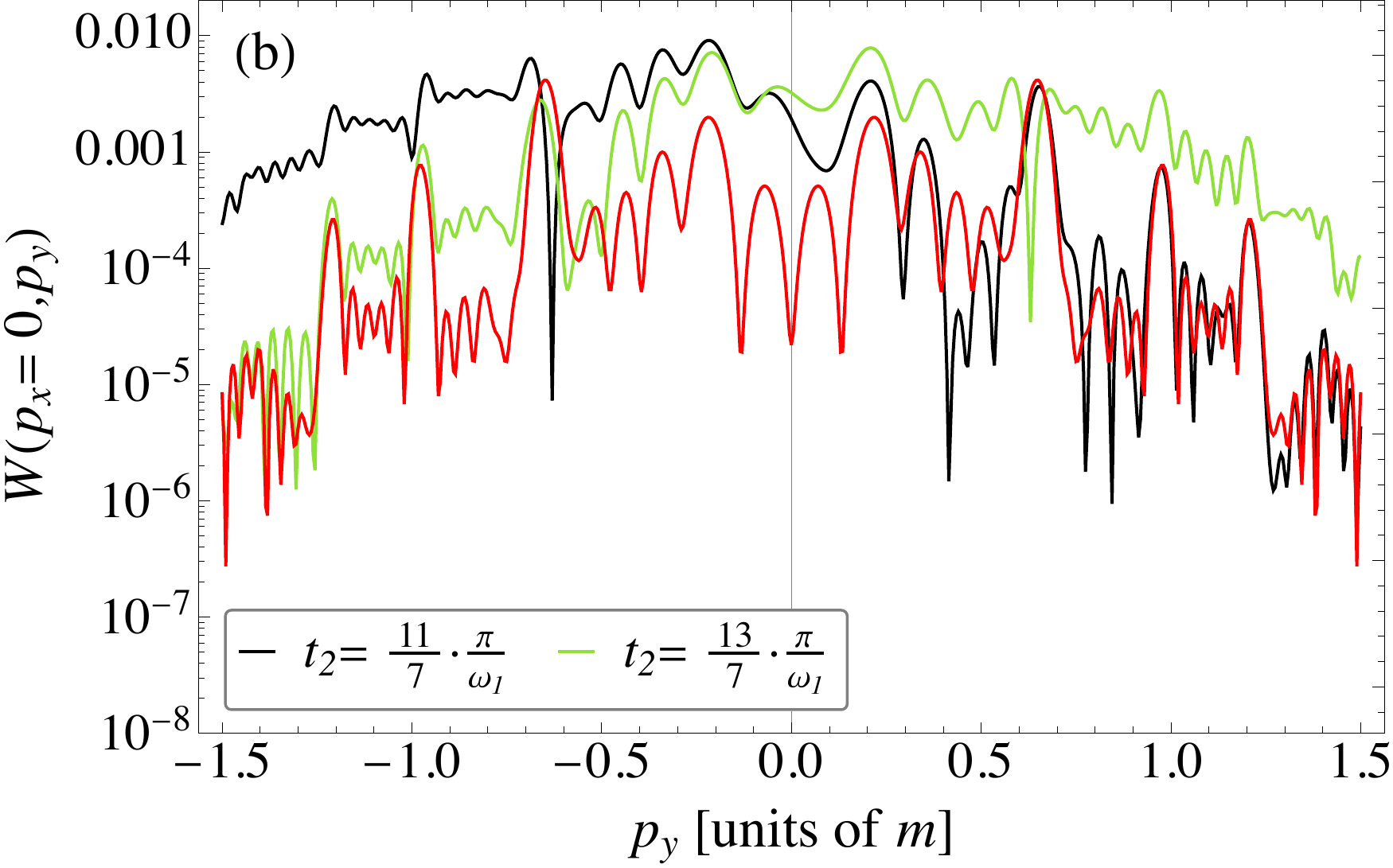}
\includegraphics[width=0.4\textwidth]{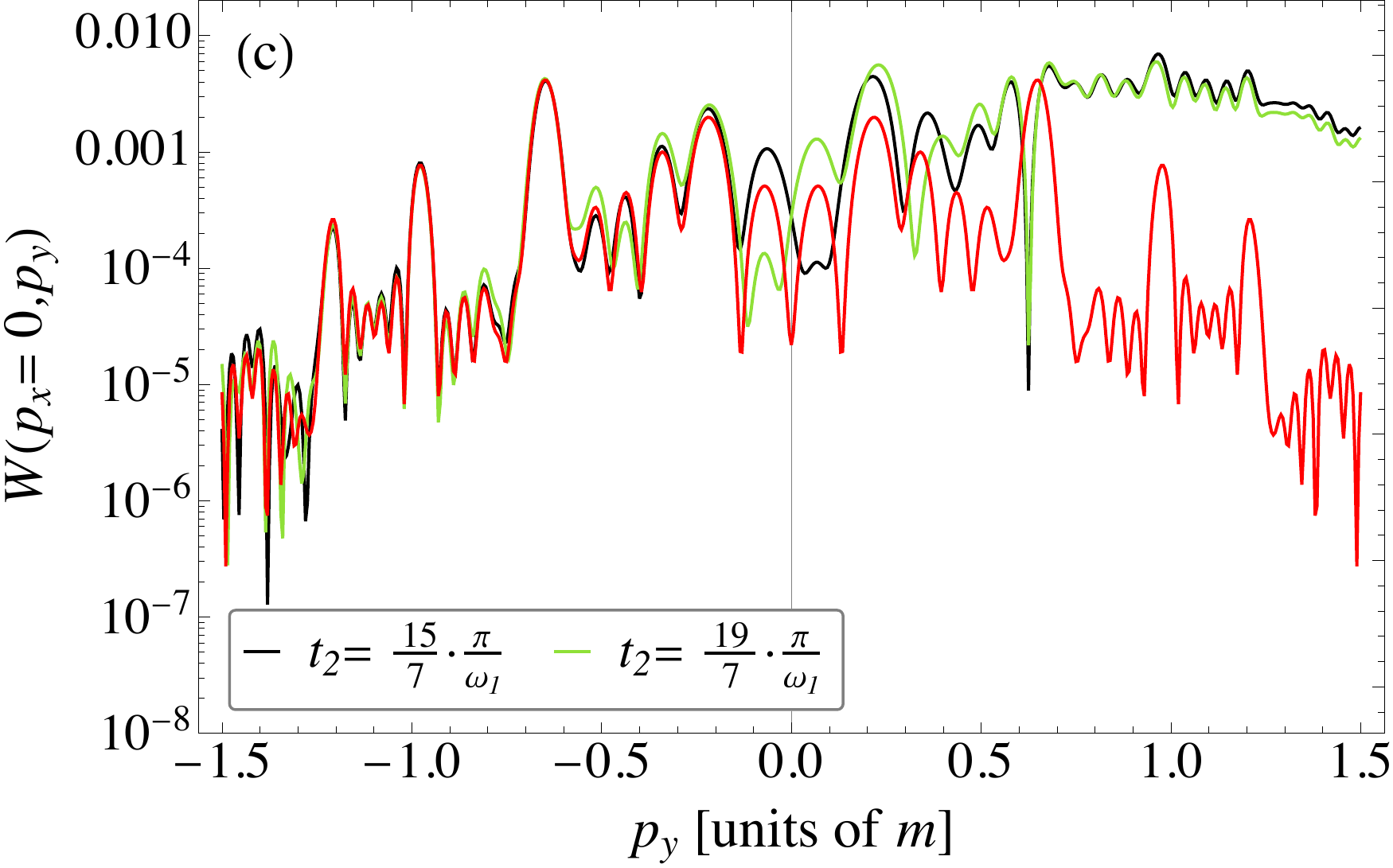}
\end{center}
\vspace{-0.5cm} 
\caption{Longitudinal momentum distribution of electrons created in a bifrequent electric field with $\xi_1=1$, $\xi_2=0.05$, $\omega_1=0.3m$, $\omega_2=1.24385m$, $N_1=6$, and $N_2=1$. Various time delays $t_2$ of the ultrashort pulse are shown by the black and green solid lines, as indicated. For comparison, the red solid curves display the distribution that results when only the main pulse is present whereas $A_2\equiv 0$.}
\label{fig-T2varies}
\end{figure}

When the second pulse is very short, the question arises to which extent the pair production is influenced by a relative delay $t_2$ between the pulses, i.e. by the precise position of the short high-frequency pulse $A_2(t)$ relative to the rather long and low-frequency pulse. 
Figure~\ref{fig-T2varies} shows the longitudinal momentum distributions of electrons created for different time delays, $\frac{\pi}{\omega_1}\le t_2 < \frac{3\pi}{\omega_1}$, within a full cycle of the main pulse; the other parameters are $\xi_1=1$, $\omega_1=0.3m$, $N_1=6$ and $\xi_2=0.05$, $\omega_2=1.24385m$ and $N_2=1$. Note that $T_2$ amounts to approximately a quarter of a cycle of the main pulse. When the short pulse is located in a region where $eA_1(t)$ is negative, which holds in Fig.~\ref{fig-T2varies}\,(a), the production of electrons with $p_y<0$ is strongly influenced and enhanced by the presence of the short pulse, whereas the spectral domain with $p_y>0$ follows very closely the momentum distribution arising when only the field $A_1(t)$ is present (shown by the red curve). This implies that the dynamical assistance is effective essentially solely for electrons with $p_y<0$, while electrons with $p_y>0$ are mainly produced by the first pulse alone. The opposite occurs when the assisting pulse lies---predominantly---in a region where $eA_1(t)>0$, corresponding to Fig.~\ref{fig-T2varies}\,(c). The cross over between these cases occurs around $t_2 = \frac{2\pi}{\omega_1}-\frac{1}{2}T_2 \approx \frac{12}{7}\frac{\pi}{\omega_1}$ where the short pulse lies symmetrically around a zero-crossing of $A_1(t)$. Figure~\ref{fig-T2varies}\,(b) refers to time delays slightly below and above this transition point. The left-right asymmetry is still pronounced in these cases, but the enhancement effect is less distinctive.

Our discussion has revealed that the momentum distributions in the combined fields encode time information: When lying close to a minimum of $eA_1(t)$, the ultrashort pulse mainly produces electrons with $p_y<0$; those electrons therefore originate predominantly from the very short time interval of the pulse $A_2(t)$. The situation is reversed when the short pulse acts during times when the long pulse runs through a maximum of $eA_1(t)$: Then it strongly amplifies the creation of electrons with $p_y>0$, but has almost no impact on the creation of electrons with $p_y<0$.   

\begin{figure}[b]
\begin{center}
\includegraphics[width=0.45\textwidth]{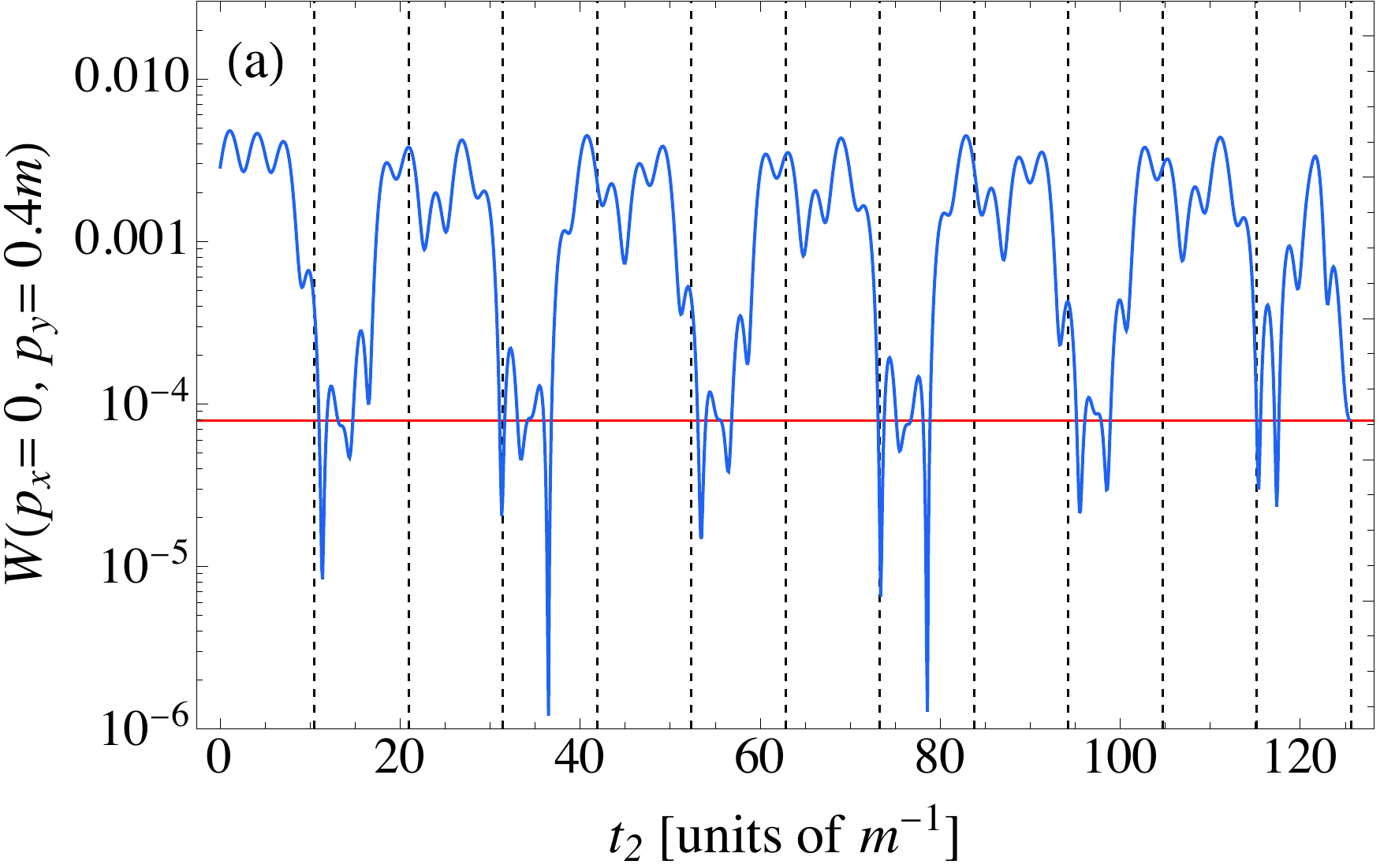}
\includegraphics[width=0.45\textwidth]{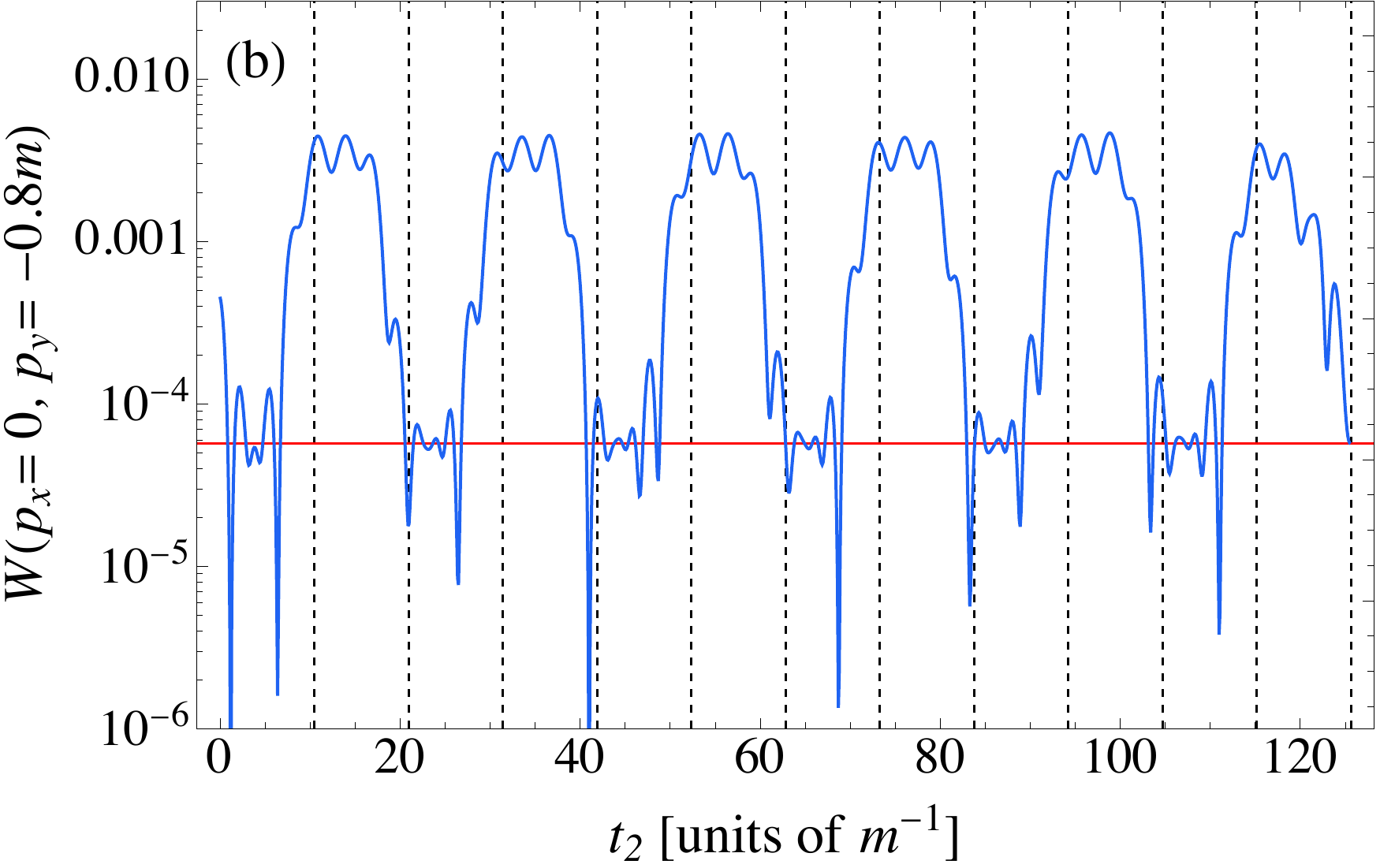}
\end{center}
\vspace{-0.5cm}
\caption{Pair production probability for fixed momenta of (a) $p_x=0$, $p_y=0.4m$ and (b) $p_x=0$, $p_y=-0.8m$ (solid curves), as function of the relative delay $t_2$,  in a bifrequent electric field with $\xi_1=1$, $\xi_2=0.05$, $\omega_1=0.3m$, $\omega_2=1.24385m$, $N_1=6$, and $N_2=1$. The vertical dashed lines indicate multiples of $\frac{\pi}{\omega_1}$, i.e. of half the period of the main pulse. The horizontal solid lines mark the respective production probabilities when solely the main pulse $A_1$ is active.}
\label{fig-delay-T2}
\end{figure}

When the time delay $t_2$ is varied continuously, an interesting periodicity appears, as Fig.~\ref{fig-delay-T2} illustrates. In panel (a) it shows the pair production probability for fixed electron momenta of $p_x=0$, $p_y=0.4m$. We see five main peak regions between $t_2\approx 10m^{-1}$ and $t_2\approx 110m^{-1}$ that correspond to the five plateau cycles of the main pulse $A_1$. (At the left and right borders, additional peaks with irregular structure appear that are associated with the turn-on and turn-off segments.) These five main peaks possess a rather complex substructure that, interestingly, is nearly identical for the first, third and fifth of them; also the second and fourth peak resemble each other closely. As the dashed vertical lines indicate half periods of the main pulse $A_1$, we see again that the pair production is enhanced when the assisting pulse $A_2$ is located close to a maximum of $eA_1$ (because the considered value of $p_y$ is positive here). Panel (b) displays a complementary situation with $p_x=0$ and $p_y=-0.8m$. Due to the negative value of $p_y$, enhancements now occur for those time delays where $A_2$ lies near a minimum of $eA_1$, leading to a horizontal shift of the peak regions by $\frac{\pi}{\omega_1}\approx 10m^{-1}$ as compared with panel (a). The substructure of the main peaks is somewhat more regular than in panel (a). In between the regions of enhancement, the production probability is often comparable with the corresponding probability when the assisting pulse is absent [red horizontal line].

\begin{figure}[t]
\vspace{0.25cm}
\begin{center}
\includegraphics[width=0.45\textwidth]{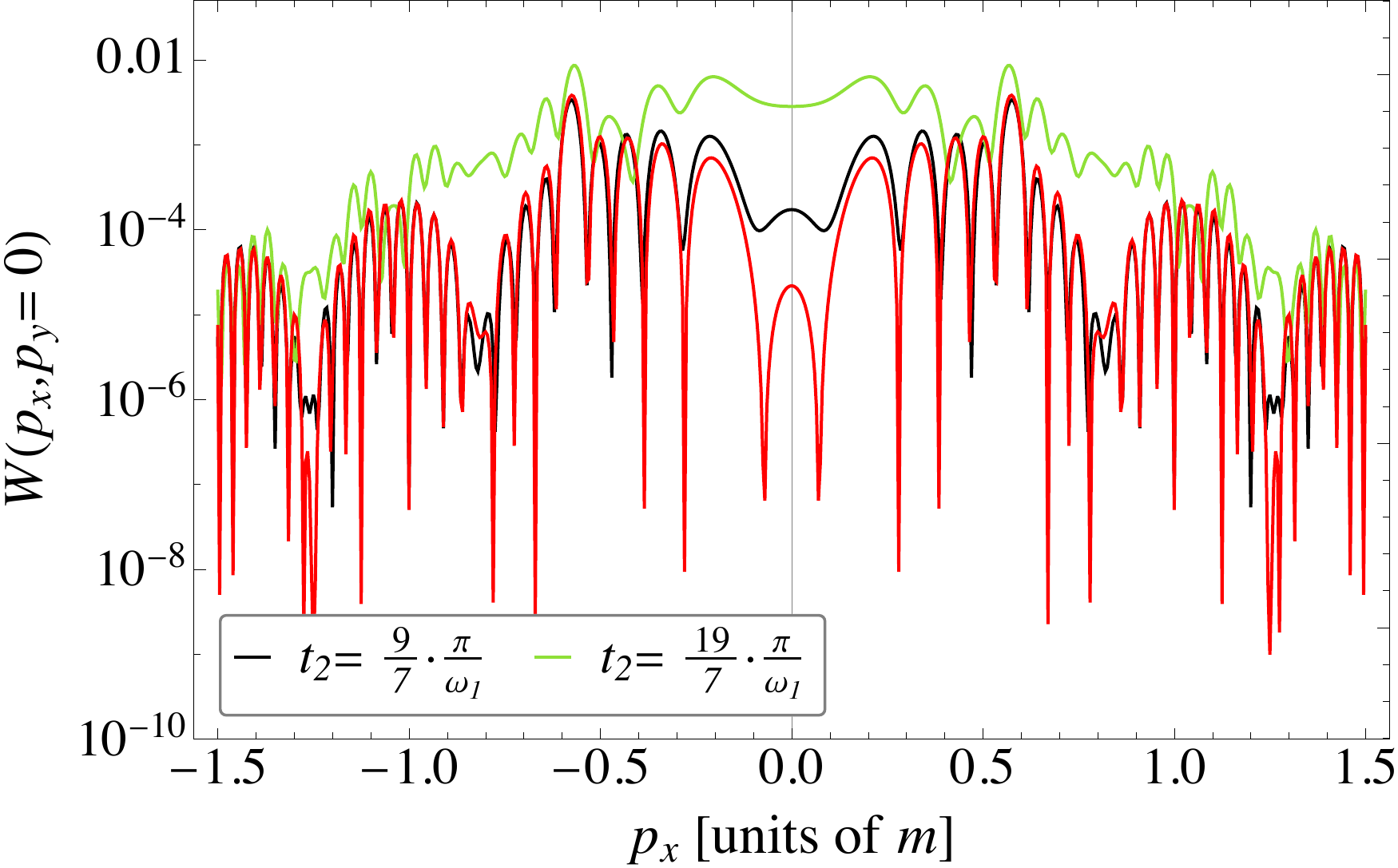}
\end{center}
\vspace{-0.5cm} 
\caption{Same as Fig.~\ref{fig-T2varies}, but for the transverse momentum,
while the longitudinal momentum is zero.}
\label{fig-T2varies-px}
\end{figure}

Variation of the time delay exerts a strong impact on the transverse momentum distribution as well. In this case, however, the spectrum is symmetric under $p_x\to -p_x$ and the difference arises between the positioning of the short pulse close to either an extremum or a zero crossing of $A_1(t)$. This is depicted in Fig.~\ref{fig-T2varies-px} for $t_2=\frac{9\pi}{7\omega_1}$ where $eA_1(t)$ runs through a minimum and $t_2=\frac{19\pi}{7\omega_1}$ where $A_1(t)\approx 0$ is close to a zero crossing. In the latter case, the short pulse has only very little effect on the pair production. When the short pulse is placed instead close to a minimum (or maximum) of $A_1(t)$ it leads to a very substantial enhancement of the particle yield.

\section{Conclusion}
Electron-positron pair production in the superposition of two oscillating electric
field pulses has been studied. The pulses were assumed to possess largely different
frequencies and durations, so that their relative positioning plays a role, which 
has turned out to be crucial.

In a first scenario, a background field of very low frequency was superimposed onto
a strong and short main pulse of high frequency that is driving the pair production. 
It was shown that the background field can modify the longitudinal momentum spectrum
of created particles in a characteristic manner. When the main pulse is located in 
the region of a minimum (maximum) of the background vector potential, the longitudinal 
electron momenta are shifted into positive (negative) direction by a corresponding amount; 
the opposite holds for the created positrons. This effect can be exploited for streak 
imaging of the background potential. We have also shown that, under suitable conditions, 
application of a background field allows to obtain information on the time intervals 
when particles with certain momenta are created predominantly.

In a second scenario, a weak ultrashort pulse of very high frequency was superimposed
onto a strong main pulse. In such a configuration, the pair production yield is known
to be enhanced by the mechanism of dynamical assistance. However, which part of the
momentum spectrum experiences the strongest enhancement was shown to depend on the
duration and relative positioning of the assisting ultrashort pulse. If it is situated
on a minimum (maximum) of the main pulse vector potential, the creation of electrons
with positive (negative) longitudinal momenta is largely enhanced. Enhancement effects
have also been found in the transverse momentum distribution when the assisting pulse
is put on a maximum or minimum of the main pulse vector potential. Also the dynamical 
enhancement caused by an ultrashort pulse can therefore be exploited to infer 
at which times particles with certain momenta are mainly created.

\section*{Acknowledgment}
This work has been funded by the Deutsche Forschungsgemeinschaft (DFG) 
under Grant No. 392856280 within the Research Unit FOR 2783/1.
N.~F. and J.~P. contributed equally to the present paper.



\begin{thebibliography}{33}
\bibitem{q0} The quasi-energy of the created positron, which formally follows from Eq.~\eqref{epsilon} by the replacements $\vec p\to-\vec p$ and $e\to-e$, attains the same value.
\bibitem{splitting} In order to split the main pulse in the middle, we ramp down the last half cycle of the first half of the pulse by a decreasing sin$^2$ window function. The first half cycle of the second half of the pulse is, accordingly, switched on by an increasing sin$^2$ window function.
\bibitem{mean-p} We note that the mean value $\langle p_y\rangle$ of 
the longitudinal momentum also reflects the shape of $eA_2$ very closely.
\bibitem{sign} We note moreover that the applied vector potential differs by an overall sign as compared with Ref.~\cite{Akal}.
\end{thebibliography}
\end{document}